\documentclass[]{aastex63}
\usepackage{CJK}
\usepackage{amsmath}
\usepackage{hyperref}

\received{...}
\revised{...}
\accepted{...}

\shorttitle{GP on BBH Mass Function}
\shortauthors{Li et al.}

\begin{document}
\begin{CJK*}{UTF8}{gbsn}

\title{A flexible Gaussian process reconstruction method and the mass function of the coalescing binary black hole systems}

\author[0000-0001-5087-9613]{Yin-Jie Li (李银杰)}
\affiliation{Key Laboratory of Dark Matter and Space Astronomy, Purple Mountain Observatory, Chinese Academy of Sciences, Nanjing 210023, Peoples Republic of China}
\affiliation{School of Astronomy and Space Science, University of Science and Technology of China, Hefei, Anhui 230026, Peoples Republic of China}

\author[0000-0001-9626-9319]{Yuan-Zhu Wang (王远瞩)}
\affiliation{Key Laboratory of Dark Matter and Space Astronomy, Purple Mountain Observatory, Chinese Academy of Sciences, Nanjing 210023, Peoples Republic of China}

\author[0000-0001-9034-0866]{Ming-Zhe Han (韩明哲)}
\affiliation{Key Laboratory of Dark Matter and Space Astronomy, Purple Mountain Observatory, Chinese Academy of Sciences, Nanjing 210023, Peoples Republic of China}
\affiliation{School of Astronomy and Space Science, University of Science and Technology of China, Hefei, Anhui 230026, Peoples Republic of China}

\author[0000-0001-9120-7733]{Shao-Peng Tang (唐少鹏)}
\affiliation{Key Laboratory of Dark Matter and Space Astronomy, Purple Mountain Observatory, Chinese Academy of Sciences, Nanjing 210023, Peoples Republic of China}
\affiliation{School of Astronomy and Space Science, University of Science and Technology of China, Hefei, Anhui 230026, Peoples Republic of China}

\author[0000-0003-4891-3186]{Qiang Yuan (袁强)}
\affiliation{Key Laboratory of Dark Matter and Space Astronomy, Purple Mountain Observatory, Chinese Academy of Sciences, Nanjing 210023, Peoples Republic of China}
\affiliation{School of Astronomy and Space Science, University of Science and Technology of China, Hefei, Anhui 230026, Peoples Republic of China}

\author[0000-0002-8966-6911]{Yi-Zhong Fan (范一中)}
\affiliation{Key Laboratory of Dark Matter and Space Astronomy, Purple Mountain Observatory, Chinese Academy of Sciences, Nanjing 210023, Peoples Republic of China}
\affiliation{School of Astronomy and Space Science, University of Science and Technology of China, Hefei, Anhui 230026, Peoples Republic of China}
\email{The corresponding author: yzfan@pmo.ac.cn (Y.Z.F)}

\author[0000-0002-9758-5476]{Da-Ming Wei (韦大明)}
\affiliation{Key Laboratory of Dark Matter and Space Astronomy, Purple Mountain Observatory, Chinese Academy of Sciences, Nanjing 210023, Peoples Republic of China}
\affiliation{School of Astronomy and Space Science, University of Science and Technology of China, Hefei, Anhui 230026, Peoples Republic of China}

\begin{abstract}

We develop a new method based on Gaussian process to reconstruct the mass distribution of binary black holes (BBHs). Instead of prespecifying the formalisms of mass distribution, we introduce a more flexible and nonparametric model with which the distribution can be mainly determined by the observed data. We first test our method with simulated data, and find that it can well recover the injected distribution. Then we apply this method to analyze the data of BBHs' observations from LIGO-Virgo Gravitational-Wave Transient Catalog 2. By reconstructing the chirp mass distribution, we find that there is a peak or a platform locating at $20-30\,M_{\odot}$ rather than a single-power-law-like decrease from low mass to high mass. Moreover, one or two peaks in the chirp mass range of $\mathcal{M}<20\,M_{\odot}$ may be favored by the data. Assuming a mass-independent mass ratio distribution of $p(q)\propto q^{1.4}$, we further obtain a distribution of primary mass, and find that there is a feature locating in the range of $(30, 40)\,M_{\odot}$, which can be related to \textsc{Broken Power Law} and \textsc{Power Law + Peak} distributions described in \cite{2020arXiv201014533T}. Besides, the merger rate of BBHs is estimated to $\mathcal{R} = 26.29^{+14.21}_{-8.96}~{\rm Gpc^{-3}~yr^{-1}}$ supposing there is no redshift evolution. 

\end{abstract}

\section{Introduction} \label{sec:intro}
The third observing run (O3) of Advanced LIGO/Virgo has already been accomplished, lasting about one year from April 2019 to March 2020, and data from the first half of the third observing run (O3a) have been released in LIGO-Virgo Gravitational-Wave Transient Catalog 2 (GWTC-2)\footnote{\url{https://www.gw-openscience.org/eventapi/html/GWTC-2/}}\citep{2020arXiv201014527A}. Together with the events observed in the first two observing runs (O1 and O2) \citep{2019PhRvX...9c1040A}, the whole catalog consists of about 50 confident gravitational wave (GW) signals from compact binary coalescences. In addition to the first detected binary black hole (BBH) merger \citep[GW150914,][]{2016PhRvL.116f1102A} and binary neutron star merger \citep[GW170817,][]{2017PhRvL.119p1101A, 2019PhRvX...9a1001A}, there are some interesting candidates (e.g., GW190426\_152155, GW190425 or GW190814) that may be originated from the neutron star-black hole mergers \citep{2020arXiv201014527A,2020arXiv201204978L,2020ApJ...891L...5H,2020ApJ...896L..44A}. The observations of very massive binaries such as GW190521 \citep{2020PhRvL.125j1102A} and the extreme mass-ratio event GW190814 have challenged the theories of stellar formation and evolution.

Investigating the population of a growing number of BBH events plays an important role in revealing the physical mechanisms associated to the formation/evolution of the stars. Some evolutionary channels may produce BBH mergers that can be observed by Advanced LIGO and Virgo, like common envelope evolution \citep{1998A&A...332..173P}, 
chemically homogeneous evolution \citep{2016A&A...588A..50M}, 
isolated binary evolution via the remnants of Population \uppercase\expandafter{\romannumeral3} stars \citep{2017MNRAS.468.5020I},
and dynamical formation \citep{1993Natur.364..421K}.
Several parameterization models have been proposed and applied to reconstruct the mass distribution of BBH events \citep{2019ApJ...882L..24A, 2020arXiv201014533T}. They found that the primary mass distribution is not a single power law with an abrupt cut-off, and there must be a feature located at $\sim 37\,M_{\odot}$. 

Is there any additional features in the mass spectrum of BBHs? Since we do not know the formation channels of BBHs very well, it may be hard to characterize the mass function of BBHs exactly.
Therefore, we propose a new nonparametric method and obtain model-independent results that are fully faithful to the basic observation data. This nonparametric method is based on Gaussian processes (GPs), which is used to reconstruct the chirp mass distribution and estimate the merger rate of BBHs' population. Our method is distinct from the widely used hierarchical formalism modeling the population properties of BBHs with parametric functions \citep{2019PASA...36...10T, 2020arXiv201014533T, 2020arXiv201104502T, 2021ApJ...913...42W, 2020ApJ...891...76D}. Since the chirp mass, which determines the leading-order gravitational wave emission during the inspiral, is much better measured than the other intrinsic parameters, we here mainly reconstruct the chirp mass distribution, and transform it to the distribution of component mass (primary mass) by assuming a fixed mass ratio distribution of $p(q)\propto q^{1.4}$ \citep{2020arXiv201014533T}. Though physical parameters (i.e., component masses and the spins) of BBHs directly relate to the formation/evolution process of their progenitor stars, their measurements always degenerate with each other when performing the parameter estimation on an individual event. Thus modeling the population properties of chirp mass is more robust \citep{2020arXiv201104502T}.

\section{Method} \label{sec:method}
\subsection{Gaussian processes} \label{model}
Gaussian processes are useful tools in GW astronomy and have been implemented in many works, such as the modeling of gravitational waveforms \citep{2017PhRvD..96l3011D,PhysRevD.97.024031} and electromagnetic counterpart light curves \citep{10.1093/mnras/sty2174},
the optimization of parameter estimation strategies \citep{2014PhRvL.113y1101M,2016PhRvD..93f4001M,2018arXiv180510457L}, 
the investigation of the binary stellar evolution \citep{2018PhRvD..98h3017T}, 
and the inference of equation of state of neutron star matter \citep{2019PhRvD..99h4049L}.
For the first time, we introduce this approach to investigate the black hole mass functions.

We model the population properties of BBHs' chirp masses in the range of $(4.5, 87)\,M_{\odot}$. The lower bound ($4.5\,M_{\odot}$) is a conservative choice, which is lower than the minimum posterior sample of the lightest BBH observation \citep{2020arXiv201014527A}, while the upper bound ($87\,M_{\odot}$) corresponds to an equal mass system with component masses of $100\,M_{\odot}$. Then this target range is divided into 30 bins, which have equal length in the logarithm space. We find that 30 bins are reasonably enough to recover the chirp mass distribution of BBHs using current observation data. We have analyzed the data with 50 bins and find that the improvement is minor, i.e., the results of 30 bins and 50 bins are highly consistent with each other. Thus the distribution of chirp mass $\mathcal{M}$ can be given by $\boldsymbol{r}=(r_1,\cdots,r_{\rm N_{\rm bin}})$, where $r_i$ is the merger rate per unit mass for the BBHs in the $i$-th bin, and $\rm N_{\rm bin}$ is the number of bins ($\rm N_{\rm bin}=30$). In order to avoid the fluctuation in each bin that caused by the currently limited number of observations, a flexible GP to model the distribution of chirp masses is 

\begin{equation}
\ln{\boldsymbol{r}}=
\begin{bmatrix}
\ln{r_1} \\
\ln{r_2} \\
\vdots  \\
\ln r_{\rm N_{\rm bin}} \\
\end{bmatrix}
\sim \mathcal{N}(\boldsymbol{\mu}, \boldsymbol{\Sigma}),
\end{equation}

where the elements of the covariance matrix $\boldsymbol{\Sigma}$ are determined by a covariance kernel, i.e., $\Sigma_{ij} = \mathcal{K}(x_i,x_j) = \rm {Cov}(\ln\lambda_i, \ln\lambda_j)$. As for the choice of the kernel type, we take the \textit{Matern-3/2} Kernel\footnote{\url{https://gpflow.readthedocs.io/en/master/notebooks/advanced/kernels.html}} $K_{\rm Matern-3/2}(\delta{x},\,l)$ \citep{2006gpml.book.....R} as the covariance kernel in this work, and use the \textit{gpflow}\citep{GPflow2017,GPflow2020multioutput} to calculate the covariance matrix. We have also tested other kernels, such as the squared-exponential kernel, but they are not better than the \textit{Matern-3/2} Kernel for this work\footnote{As introduced in \cite{2006gpml.book.....R}, the squared-exponential kernel provides a strong smoothness that sometimes is unrealistic for modeling some physical processes, and alternatively, \textit{Matern-3/2} kernel may be more flexible.}.
Therefore, the covariance kernel is $\mathcal{K}(x_i,x_j ) = \sigma^2K_{\rm Matern-3/2}(|{x_i-x_j}|,\,\,l)$, where $x_i$ corresponds to the median point of the $i$-th bin in logarithmic space, and the hyper-parameters $\sigma$ and $l$ determine the functions' behavior: $\sigma$ sets the amplitude of the function we modeled, while $l$ governs the length scale over which the correlations occur. Longer length scales cause long-range correlations, whereas for short length scales, function values are strongly correlated only if their respective inputs are very close to each other.

\subsection{Bayesian inference} \label{inference}
\subsubsection{Hyper-parameters}
The prior of chirp mass function $\boldsymbol{r} = (r_1,\cdots,r_{\rm N_{\rm bin}})$ is generated by the GP: $\pi(\boldsymbol{r}|\boldsymbol{\mu},\sigma,l)$. In principle, one could choose the mean values $\boldsymbol{\mu} = (\mu_1,\dots,\mu_{\rm N_{\rm bin}})$ and hyper-parameters ($\sigma$, $l$) depending on the prior information of the chirp mass distribution. 
In our implementation, we set $\boldsymbol{\mu}=\boldsymbol{0}$, which is a general option in GP. As for $\sigma$, it will potentially affect the smoothness of the function though, while its effect is weaker than that caused by length scale, thus we fix it in our inference. We find that $\sigma=5$ is appropriate to generate a prior in our Bayesian inference\footnote{If we choose a smaller one like $\sigma=2$, the prior is too narrow that will bias the results of inference; while if we choose a larger one like $\sigma=10$, the inference becomes more time-consuming.}.
Another hyper-parameter $l$ is the critical parameter for our model, which mainly shapes the function. Larger $l$ makes the model stiffer that will ignore the details, while smaller $l$ makes the model softer that shows more details but may give rise to over-fitting. For completeness, we perform several inferences with different length scales, and then compare the related evidences.

\subsubsection{Likelihood}\label{sec:llh}
The construction of the likelihood function in this work is different from the standard hierarchical Bayesian inference \citep{2012PhRvD..86l4032A, 2019PASA...36...10T}, widely used in studying the population properties of gravitational wave sources \citep{2019ApJ...882L..24A, 2020arXiv201014533T, 2020arXiv201104502T}.

Given the detection numbers in all bins $\boldsymbol{k}=(k_1,\cdots,k_{\rm N_{\rm bin}})$, the likelihood can be expressed as the function of expected detection numbers in all bins $\boldsymbol{\lambda} = (\lambda_1, \dots, \lambda_{\rm N_{\rm bin}})$, which is 
\begin{equation}
\mathcal{L}(\boldsymbol{k}|\boldsymbol{\lambda}) = \prod_{i=1}^{\rm N_{\rm bin}}{e^{-\lambda_i} \frac{{\lambda_i}^{k_i}}{{k_i}!}}.
\end{equation}
However, there are uncertainties in the measurements of chirp mass for individual event, i.e., $\boldsymbol{k}$ remains uncertain, thus we should integral the likelihood over the probability of $\boldsymbol{k}$. We define $\mathcal{C}_{\boldsymbol{k}}$ as the set of $\boldsymbol{k}$, and $P_{\boldsymbol{k}}$ as the probability distribution of $\boldsymbol{k}$, i.e., $\boldsymbol{k}\sim P_{\boldsymbol{k}}$, so the integrated likelihood becomes
\begin{equation}
\mathcal{L}(P_{\boldsymbol{k}}|\boldsymbol{\lambda}) =\sum_{\boldsymbol{k} \in \mathcal{C}_{\boldsymbol{k}}}\mathcal{L}(\boldsymbol{k}|\boldsymbol{\lambda}) P_{\boldsymbol{k}}(\boldsymbol{k})=\sum_{\boldsymbol{k} \in \mathcal{C}_{\boldsymbol{k}}} \prod_{i=1}^{\rm N_{\rm bin}}{e^{-\lambda_i} \frac{{\lambda_i}^{k_i}}{{k_i}!}}P_{\boldsymbol{k}}(\boldsymbol{k}).
\end{equation}
The likelihood of data $\{d_j\}$ given $\boldsymbol{\lambda}$ can be expressed as 
\begin{equation}\label{llh}
\mathcal{L}(\{d_j\}|\boldsymbol{\lambda}) = \sum_{\boldsymbol{k} \in \mathcal{C}_{\boldsymbol{k}}} \mathcal{L}(\boldsymbol{k}|\boldsymbol{\lambda})~\mathcal{L}(\{d_j\}|\boldsymbol{k}),
\end{equation}
where $\mathcal{L}(\{d_j\}|\boldsymbol{k})$ is the likelihood of data $\{d_j\}$ given $\boldsymbol{k}$, which can be reasonably related to $P_{\boldsymbol{k}}$ ($P_{\boldsymbol{k}}$ can be directly determined by the data of the observed events $\{d_j\}$) by $P(\boldsymbol{k})\propto\mathcal{L}(\{d_j\}|\boldsymbol{k})$. To evaluate $\mathcal{L}(\{d_j\}|\boldsymbol{\lambda})$, we use 10000 sets of samples $\{\boldsymbol{S}_{1},\dots,\boldsymbol{S}_{10000}\}$, and each sample $\boldsymbol{S}_{m} = (\mathcal{M}_1,\dots,\mathcal{M}_{\rm N_{\rm {det}}})_m$ is randomly chosen from the posterior samples of $\rm N_{\rm det}$ observed events, and for each posterior sample $\mathcal{M}_i$, the probability of being chosen is $\propto \frac{1}{\pi_{\varnothing}(\theta_i)}$, ( ${\pi_{\varnothing}(\theta_i)}$ is it's default prior). By counting the numbers of sample in each chirp mass bin for a given $\boldsymbol{S}_{m}$, we can get a corresponding $\boldsymbol{k}_m$, i.e., $\{\boldsymbol{S}_{1},\dots,\boldsymbol{S}_{10000}\}$ is converted to $\{\boldsymbol{k}_{1},\dots,\boldsymbol{k}_{10000}\}$. Thus Eq.~(\ref{llh}) can be evaluated by
\begin{equation}\label{datalam}
\mathcal{L}(\{d_j\}|\boldsymbol{\lambda}) = \sum_{\boldsymbol{k} \in{\{\boldsymbol{k}_{1},\dots,\boldsymbol{k}_{10000}\}}} \prod_{i=1}^{\rm N_{\rm bin}}{e^{-\lambda_i} \frac{{\lambda_i}^{k_i}}{{k_i}!}}.
\end{equation}

\subsection{Selection effect}\label{select}

The expectation of the detection number in the $i$-th chirp mass bin ($\lambda_i$) is a function of $r_i$,
\begin{equation}\label{Eq:sel}
\lambda_i = \lambda_i^{\rm O1O2} + \lambda_i^{\rm O3a} =\sum_{o \in (\rm{O1O2,O3a})} \int \int{}{}{T_{\rm obs}^o r_{i} P_{\rm det}^o(z | \theta) f_i(\theta) \frac{dV_{\rm c}}{dz} \frac{1}{1+z} dz d\theta} = F(r_i), 
\end{equation}
where $P_{\rm det}(z | \theta)$ is the probability that an event with parameters $\theta$ at redshift $z$ is detectable, ${dV_{\rm c}}/{dz}$ is the differential comoving volume per unit redshift, and $r_i$ is the merger rate of the BBHs with chirp masses in the $i$-th bin. Following \cite{2020arXiv201014533T}, we calculate $F(r_i)$ with injections\footnote{The injection campaign's data is now publicly available in \url{https://dcc.ligo.org/LIGO-P2000434/public}}. Using a Monte Carlo integral over the found injections, Eq. (\ref{Eq:sel}) is estimated as 
\begin{equation}
\hat{\lambda_i} = \sum_{o \in \rm{(O1O2,O3a)}} \frac{r_i~{\left \langle VT\right \rangle}}{N_{\rm{inj},\textit{i}}^o}\sum_{j=1}^{N_{\rm{found}, \textit{i}}^o} {\frac{p_i(\theta_j)}{p_{{\rm inj}, \textit{i}}(\theta_j)}},
\end{equation}
\begin{equation}\label{VT0}
{\left \langle VT\right \rangle}=\int_{0}^{2.3} {T_{obs}\frac{dV_{c}}{dz} \frac{1}{1+z} dz },
\end{equation}
where ${p_i(\theta_j)}$ is the normalized probability distribution of the assumed BBH population in $i$-th bin, and ${p_{{\rm inj},i}(\theta_j)}$ is for injection campaign. 
Since the number of bins is sufficient (i.e. the bins are narrow enough) and the mass ratio distribution of the injection campaign (i.e. a PowerLaw with index $\beta_q=2$) is close to the inferred distribution in \cite{2020arXiv201014533T} (i.e. PowerLaw with index $\beta_q=1.4^{+2.49}_{-1.47}$), we can make the assumption that the mass ratio and chirp mass in each bin share the same distribution as injection campaign. 
As for spin parameters, we use the distribution inferred by the default model in \cite{2020arXiv201014533T}.
Thus, we have ${p_i(\theta_j)}/{p_{{\rm inj},i}(\theta_j)}={(p(z_j)p(s1z_j, s2z_j))}/{(p_{{\rm inj}}(z_j)p_{\rm inj}(s1z_j,s2z_j))}$, where $p(s1z_j,s2z_j)$ and $p_{\rm inj}(s1z_j,s2z_j)$ are aligned spin distribution from \cite{2020arXiv201014533T} and that of the injection campaign, respectively, while ${p(z_j)}$ and ${p_{\rm inj}(z_j)}$ correspond respectively to the redshift distribution based on our non-evolution assumption and that of the injection campaign in which $\mathcal{R}(z) \propto (z+1)^2$ is assumed.
$N_{{\rm inj},i}$ ($N_{{\rm found},i}$) is the number of total injections (detected injections) with chirp mass located in $i$-th bin. Note that the distribution of injection campaign in \cite{2020arXiv201014533T} is modeled by component masses $m_1$ and $m_2$, but $N_{{\rm inj},i}$ depends on the distribution of chirp mass. Therefore, we firstly generate $N_{\rm inj}$\footnote{The total number of injections is $7.7 \times 10^7$ for O3a and $7.1 \times 10^7$ for O1 and O2 \citep{2020arXiv201014533T}.} injections with $(m_1, m_2)$ following \cite{2020arXiv201014533T}, and convert them to $\mathcal{M}$, then $N_{{\rm inj},i}$ is estimated by counting the number of injections with $\mathcal{M}$ in the $i$-th bin.
${\left \langle VT\right \rangle}$ is the surveyed space-time calculated by Eq.~(\ref{VT0}). 

Putting Eq.~(\ref{Eq:sel}) into Eq.~(\ref{datalam}), the likelihood of data $\{d_i\}$ given the distribution of chirp mass $\boldsymbol{r}$ can be rewritten as 
\begin{equation}\label{Eqlikelihood}
\mathcal{L}(\{d_j\}|\boldsymbol{r}) = \sum_{\boldsymbol{k} \in{\{\boldsymbol{k}_{1},\dots,\boldsymbol{k}_{10000}\}}} \prod_{i=1}^{\rm N_{\rm bin}}{e^{-F(r_i)} \frac{{F(r_i)}^{k_i}}{{k_i}!}}.
\end{equation}

\section{Reconstruction with Simulated Data}
To check the performance of our method, we carry out the reconstruction procedure with the simulated data. We assume an initial distribution for chirp mass, which we are going to recover, as
\begin{equation}\label{inj}
P(\mathcal{M}|m_{\rm min},m_{\rm max},\alpha,\mu,\sigma,\lambda_{\rm peak}) = (1-\lambda_{\rm peak})\mathcal{P}(\mathcal{M}|-\alpha,m_{\rm min},m_{\rm max})+\lambda_{\rm peak}\mathcal{G}(\mathcal{M}|\mu,\sigma),
\end{equation} 
where $\mathcal{P}(\mathcal{M}|-\alpha,m_{\rm min},m_{\rm max})$ is a normalized power-law distribution with spectral index $-\alpha$ and low-mass (high-mass) cut-off $m_{\rm min}$ ($m_{\rm max}$), $\mathcal{G}(\mathcal{M}|\mu,\sigma)$ is a normalized Gaussian distribution with the mean $\mu$ and width $\sigma$, and $\lambda_{\rm peak}$ is the fraction of Gaussian component. Here, $m_{\rm min}$, $m_{\rm max}$, $\alpha$, $\mu$, $\sigma$, and $\lambda_{\rm peak}$ are set to $5\,M_{\odot}$, $80\,M_{\odot}$, $3$, $25\,M_{\odot}$, $5\,M_{\odot}$, and $0.12$ respectively. The distribution for mass ratio follows a Power Law with $p(q) \propto q^2$ and $m_2 > 2\,M_\odot$. For simplicity, the orbit-aligned dimensionless spin parameters are fixed to zero. In the injection, we assume that the merger rate is uniformly distributed in source frame with boundary of $ max({d_{\rm L}}) = 10000 \rm Mpc$, and the sky locations and binary orientation are set isotropic. Besides, we use the waveform template {\sc IMRPhenomD} \citep{2016PhRvD..93d4006H} to generate the simulated signals, and inject them into the noise generated by the representative \textsc{Advanced LIGO Mid Noise} PSD \citep{2018LRR....21....3A}, which is appropriate for the state of LIGO in O3. We define a BBH event being `detected' when its network SNR $> 12$. As for the parameter estimation for each `detected' event, we adopt the same waveform template as injection (i.e., {\sc IMRPhenomD}), and use the user friendly GW parameter inference package \textit{Bilby} \citep{2019ApJS..241...27A}  and nested sampling sampler \textit{PyMultinest} \citep{2016ascl.soft06005B}. 

\begin{figure}
\gridline{\fig{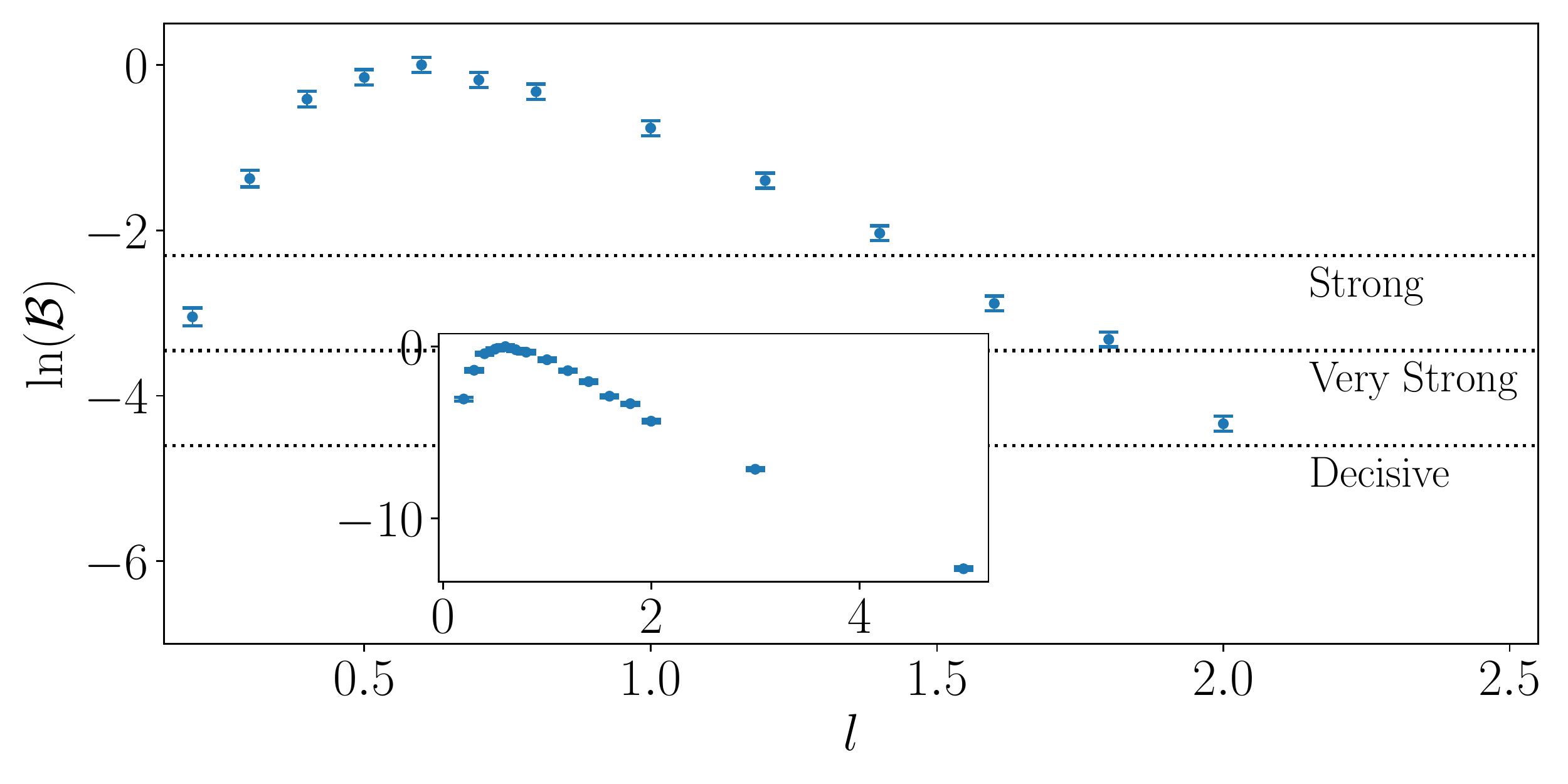}{0.48\textwidth}{}
\fig{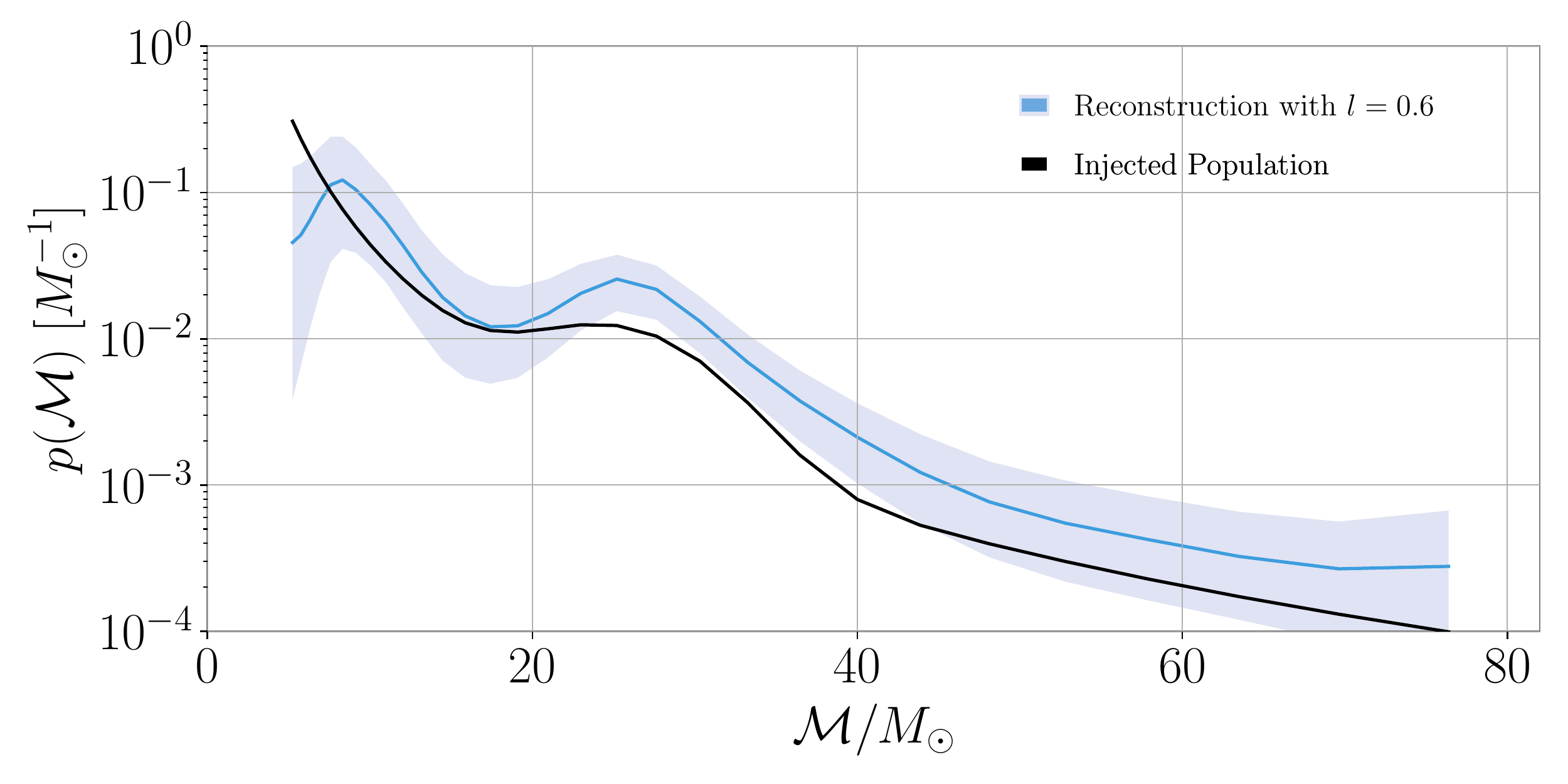}{0.5\textwidth}{}}
\gridline{\fig{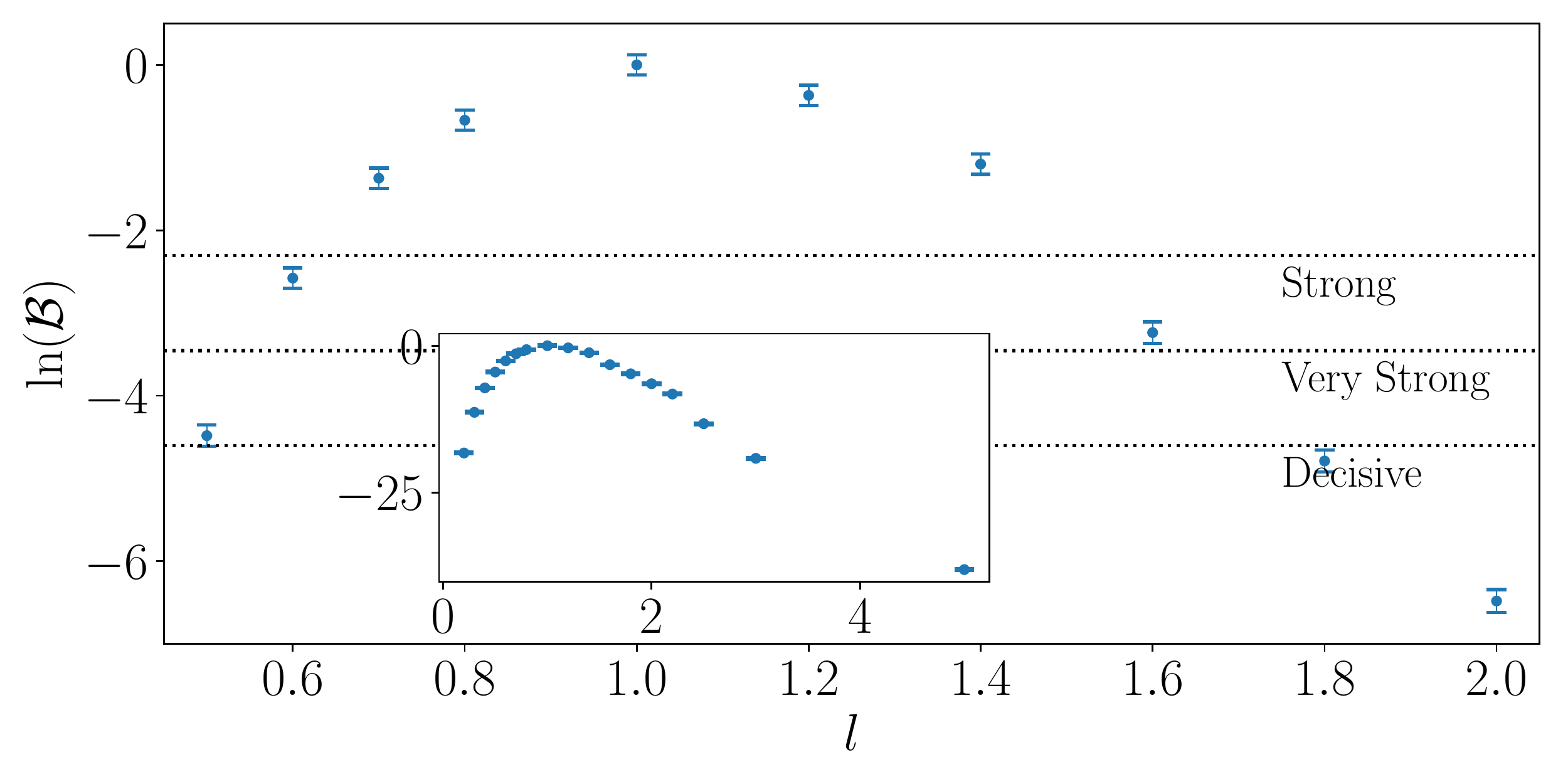}{0.48\textwidth}{}
\fig{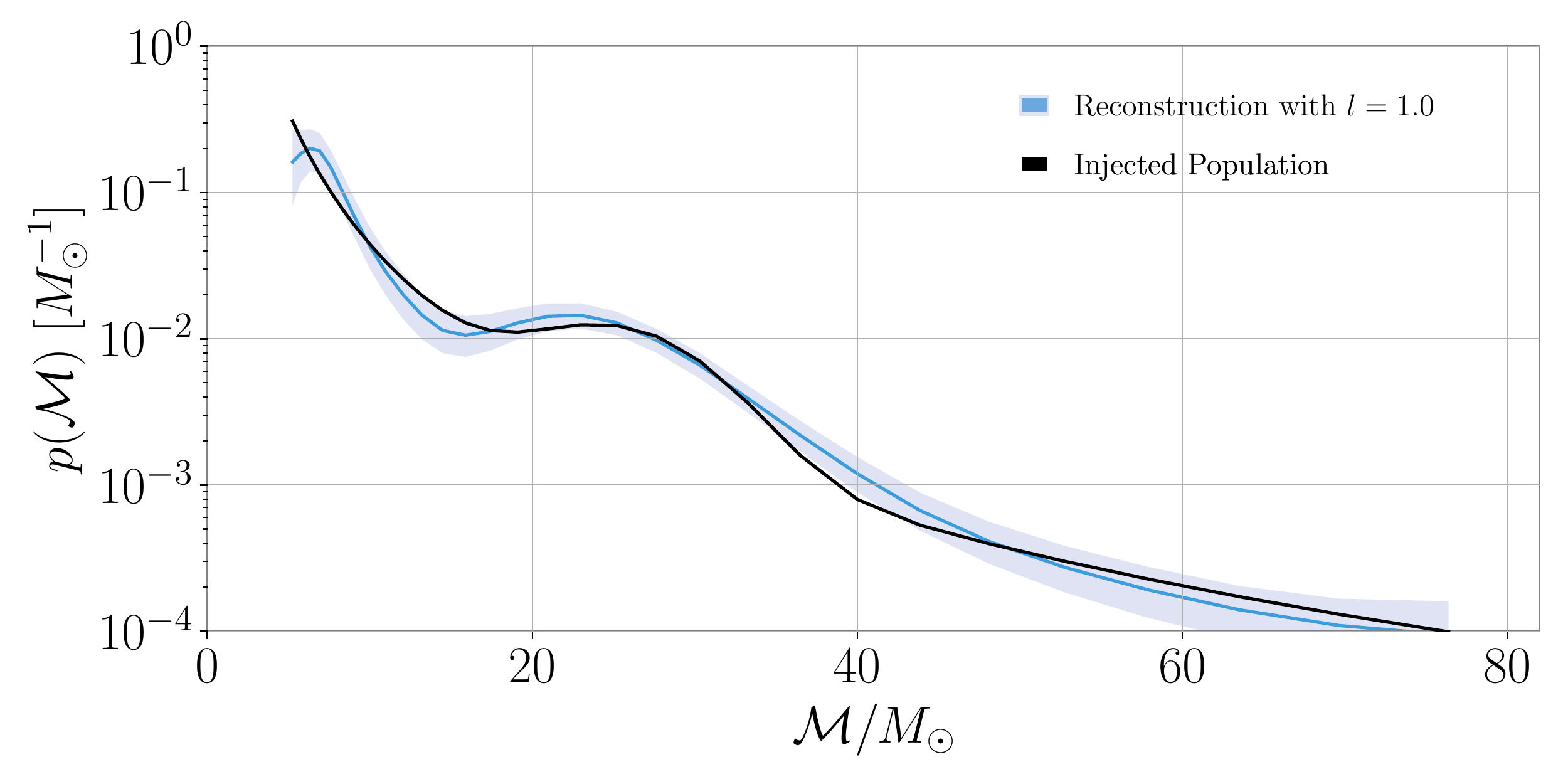}{0.5\textwidth}{}}
\gridline{\fig{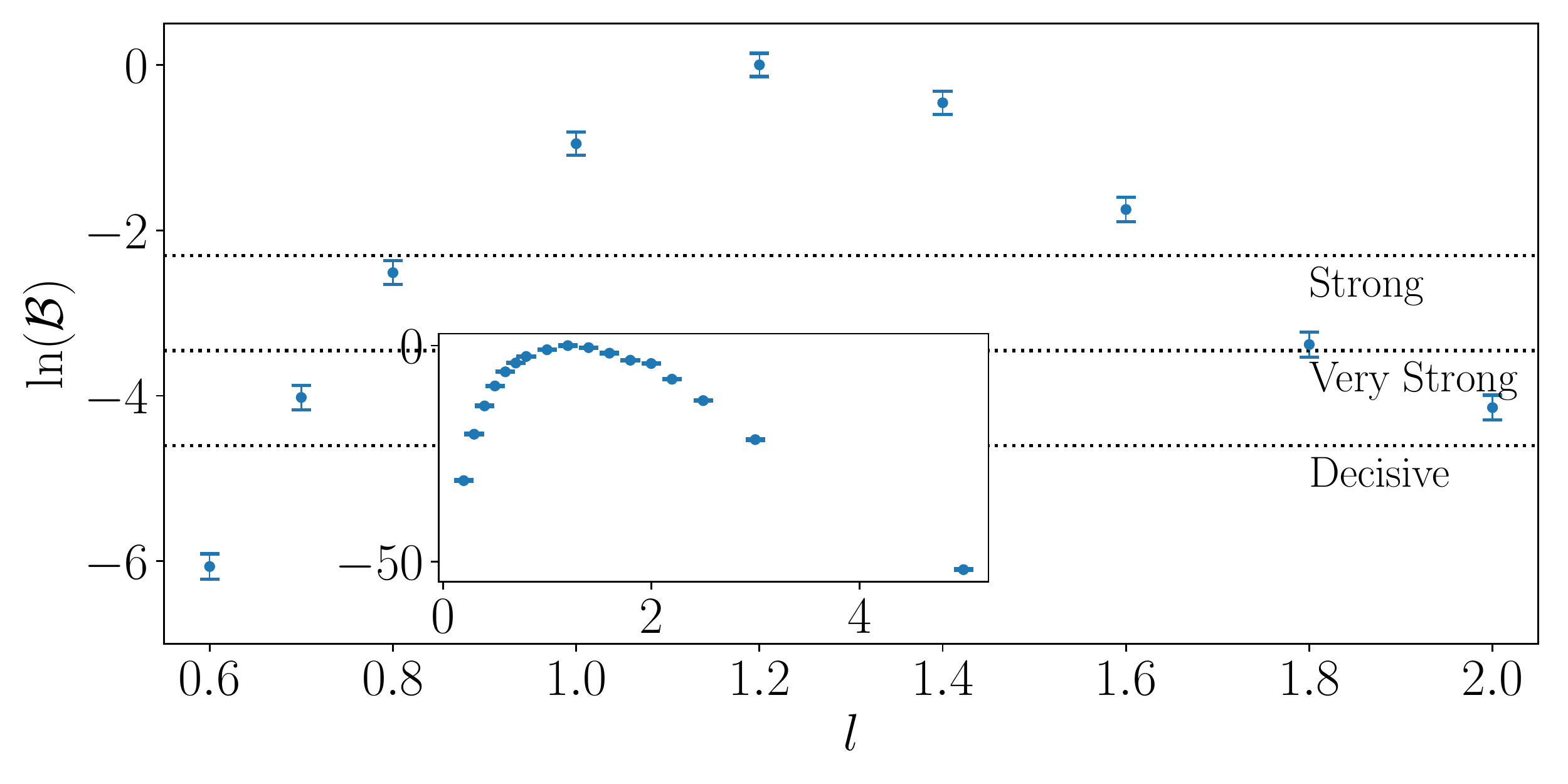}{0.48\textwidth}{}
\fig{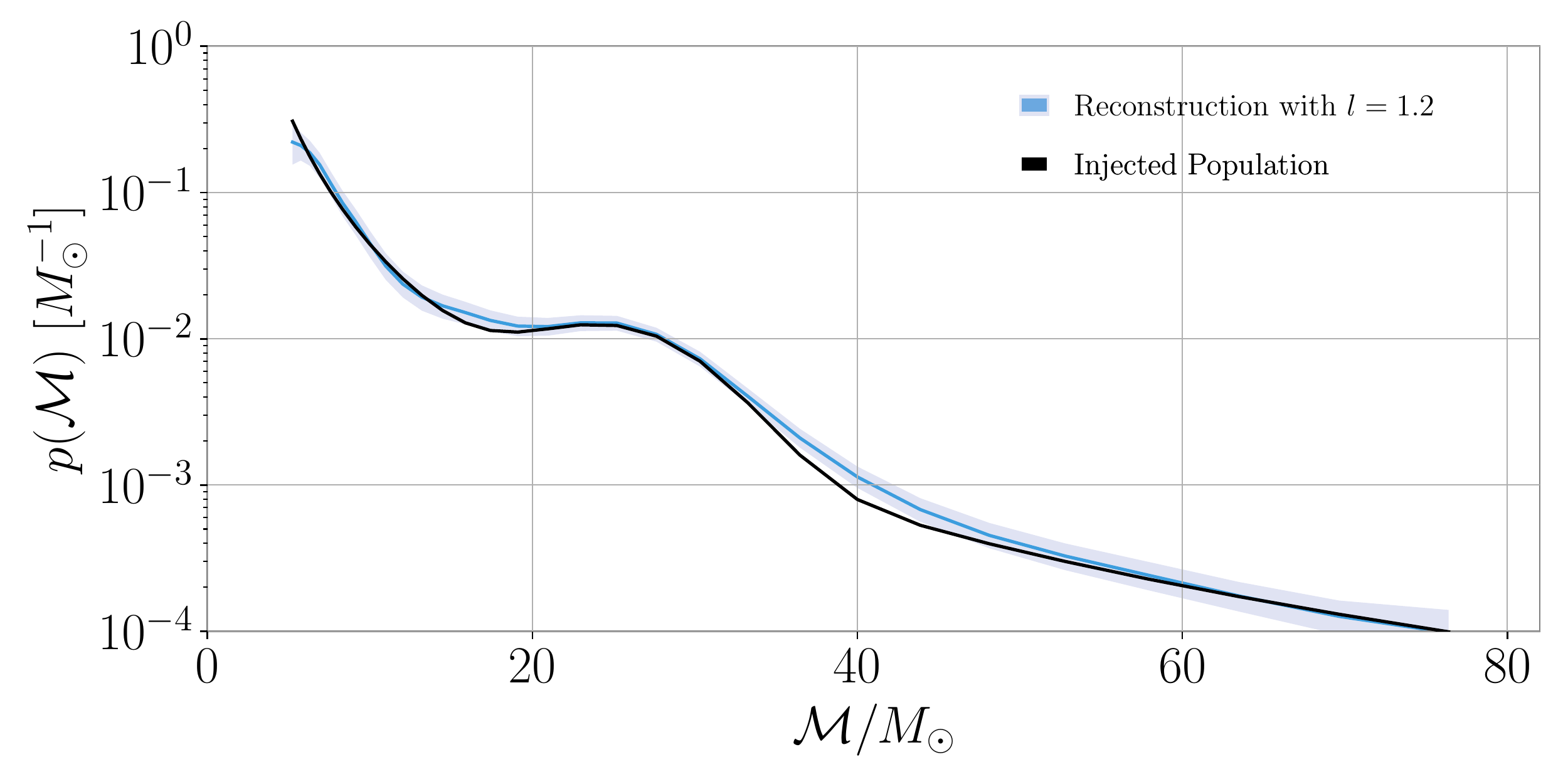}{0.5\textwidth}{}}
\caption{Left column: Bayes factors corresponding to the reconstruction of chirp mass function with different length scales upon 40 (top panel), 300 (middle panel), and 1000 (bottom panel) simulated detections, which are evaluated relative to the model with the strongest Bayesian evidence. { The insert in each figure is for a wider range of $l$.} Right column: Normalized distributions of chirp mass corresponding to the highest Bayes factor in the case of 40 (top panel), 300 (middle panel), and 1000 (bottom panel) simulated detections. The solid blue curve is the mean value over the posterior population distribution, the shaded region shows the 90\% credible interval, and the black curve represents the injected distribution. Note that the lower (upper) boundary of the mass function corresponds to the midpoint of the first (last) bin.}
\label{sim_rec}
\end{figure}

By applying the method described in Sec.~\ref{sec:method}, we reconstruct the chirp mass function in the cases of 40, 300, and 1000 simulated detections. As shown in the left column of Fig.~\ref{sim_rec}, in each case, we provide the Bayes factors of the models with different length scales, and each Bayes factor is evaluated relative to the model with the strongest Bayesian evidence. The right column of Fig.\ref{sim_rec} shows the normalized distributions of chirp mass with the most favored length scales, and we can find that when the number of detected events increases, the injected distribution is more accurately recovered. In the case of 40 detections, the distribution in high-mass range rises up, and this may be caused by the fluctuation of detections around the lower boundary\footnote{Note that the mass functions are normalized.}, which can be avoided with more detections. Generally, our method is capable of recovering the underlying mass distribution of BBHs. 

The reconstructions with other length scales are displayed in Fig.~\ref{sim_app} in the Appendix. As expected, the mass function with larger $l$ is harder to recover the detail of the underlying structure, while the mass function with shorter $l$ is overfitted and captures the fake structures produced by the fluctuation. \cite{HaroldJeffreysWrited1961Theory} pointed out that $\log_{10}{\mathcal{B}} > 1$ ($\log_{10}{\mathcal{B}} > 1.5$) can be interpreted as a strong (very strong) preference for one model over another, and $\log_{10}{\mathcal{B}}>2$ as decisive evidence. Thus, considering the mass distributions (with different length scales) and their corresponding Bayes factors, we can exclude the fake structures in the low-mass range (i.e., $ < 20\,M_{\odot}$) and determine whether the distribution that ignores the peak at $25\,M_{\odot}$ is disfavored or not, especially for the case with 300 and 1000 detections.

\section{Reconstruction with Data of GWTC-2}\label{result}
In this section, we apply our method to study the BBH population properties using the real data from GWTC-2, which includes the BBH observations during O1, O2 and O3a. Following \cite{2020arXiv201014533T}, we choose the false alarm rate (FAR) of $1\rm{yr^{-1}}$ as the threshold to select the events. To increase the purity of the sample and avoid misidentification of BBH \citep{2020ApJ...892...56T}, the GW190426, GW190719, and GW190909 are not included in this analysis. In addition, we further exclude the unusual asymmetry system GW190814 \citep{2020ApJ...896L..44A}, which is a significant outlier \citep{2020arXiv201014533T}.
For BBH events in O1 and O2, we use the `overall' posterior samples that released in \cite{2019PhRvX...9c1040A}\footnote{\url{https://dcc.ligo.org/LIGO-P1800307/public}}. While for BBH events in O3a, the `publication' posterior samples are acquired from \cite{2020arXiv201014527A}\footnote{\url{https://dcc.ligo.org/LIGO-P2000061/public}}.

\begin{figure*}
\gridline{\fig{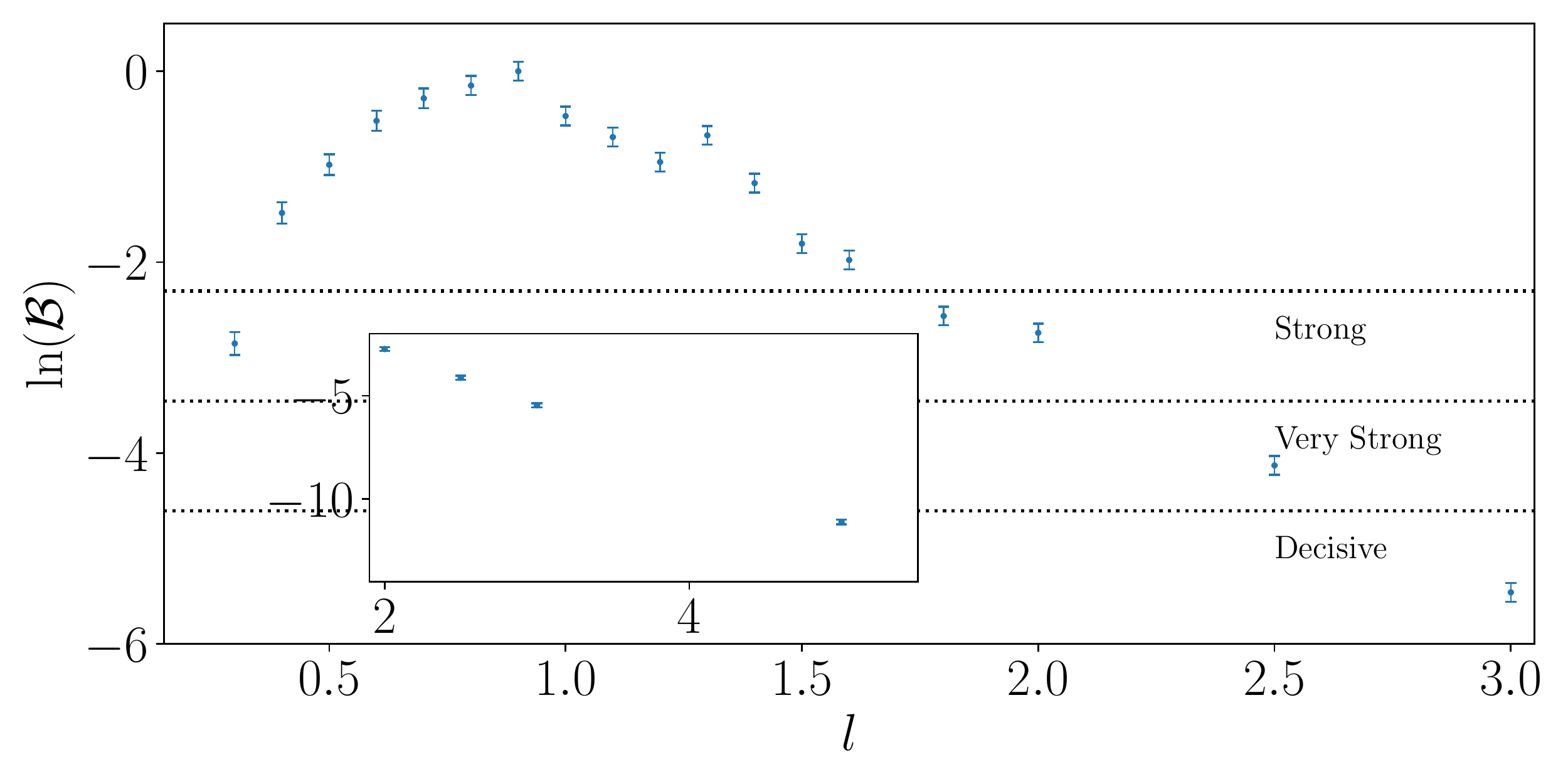}{0.5\textwidth}{}
\fig{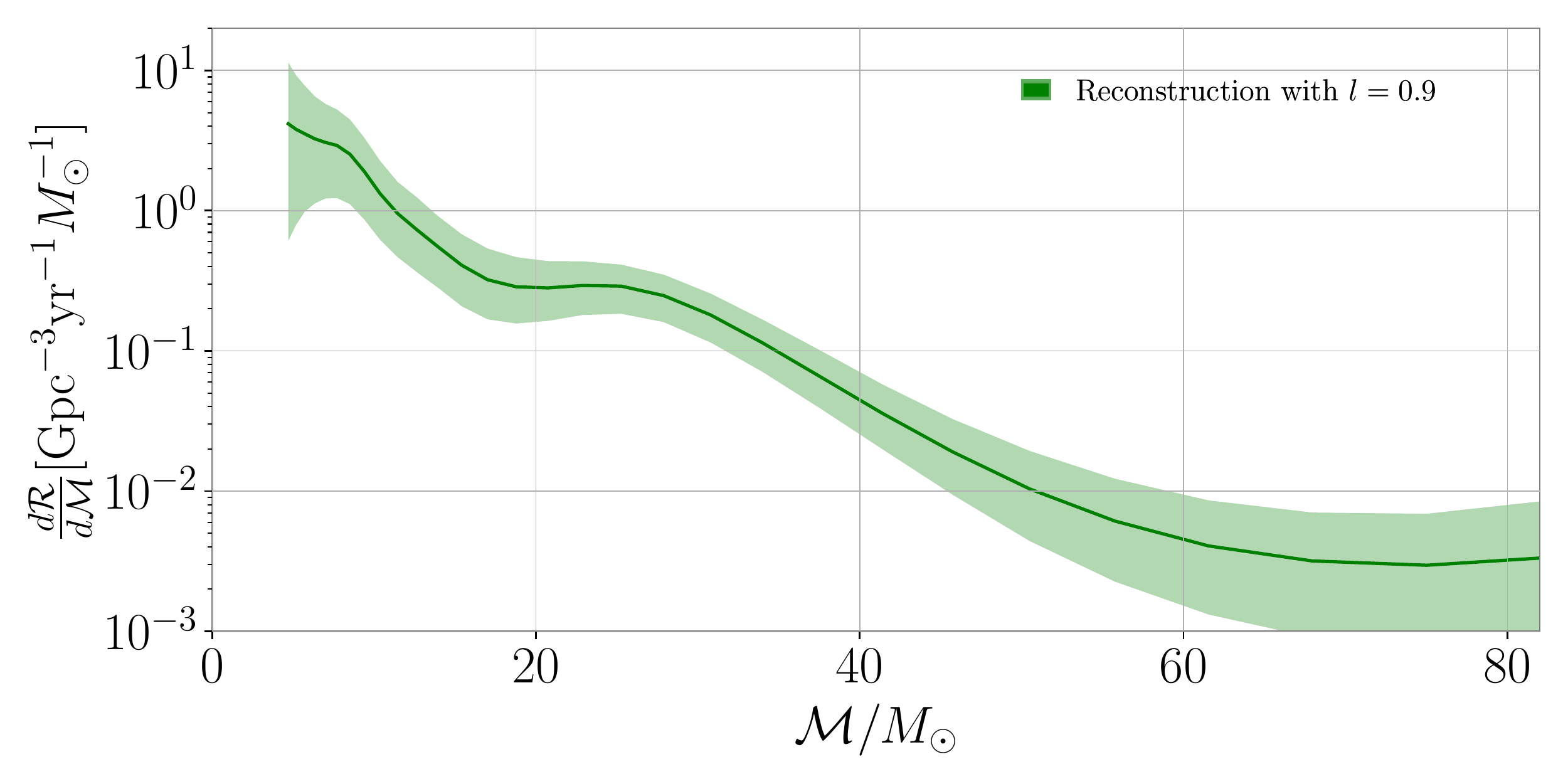}{0.5\textwidth}{}}
\caption{Refer to Fig.~\ref{sim_rec}, but for real data of detections.}
\label{Ls}
\end{figure*}

Our main results are reported in Fig.~\ref{Ls}, where we provide the Bayes factors (relative to the most favored model) of the models with different length scales (as shown in the left panels) and the most favored chirp mass distribution (with best length scale) reconstructed from the observed data (as shown in the left panels). 
Interestingly, we find that there is an additional feature located in the range of $(20,\,30)\,M_{\odot}$, and there might be a peak located in $(5,\,10)\,M_{\odot}$. Meanwhile, abrupt cut off in the high-mass range is significantly disfavored. The reconstructions with different length scales are presented in Fig.~\ref{real_app} of the Appendix \ref{length}. For the results with small length scales, like $l = 0.3$ and $l = 0.5$, there are three peaks located at $(5,\,10)\,M_{\odot}$, $(10,\,20)\,M_{\odot}$, and $(20,\,30)\,M_{\odot}$, respectively. When $1 < l < 1.5$, the two peaks in the low-mass range disappear, while the one in $(20,\,30)\,M_{\odot}$ still exists. As for the results with large length scales, like $l=5$, all additional features disappear, and the distribution becomes single-power-law-like. Based on the Bayes factors corresponding to the models with different length scales, we conclude that there must be some additional features in the chirp mass function of BBHs rather than a single-power-law-like distribution. The peak in $(20,\,30)\,M_{\odot}$ or maybe a break at $\sim 30\,M_{\odot}$, is supported by current observations, while the additional features in the range of $(5,\,10)\,M_{\odot}$ and $(10,\,20)\,M_{\odot}$ can not be confidently identified yet, which will be clear with future observations. 

\begin{figure}\label{m1}
\gridline{\fig{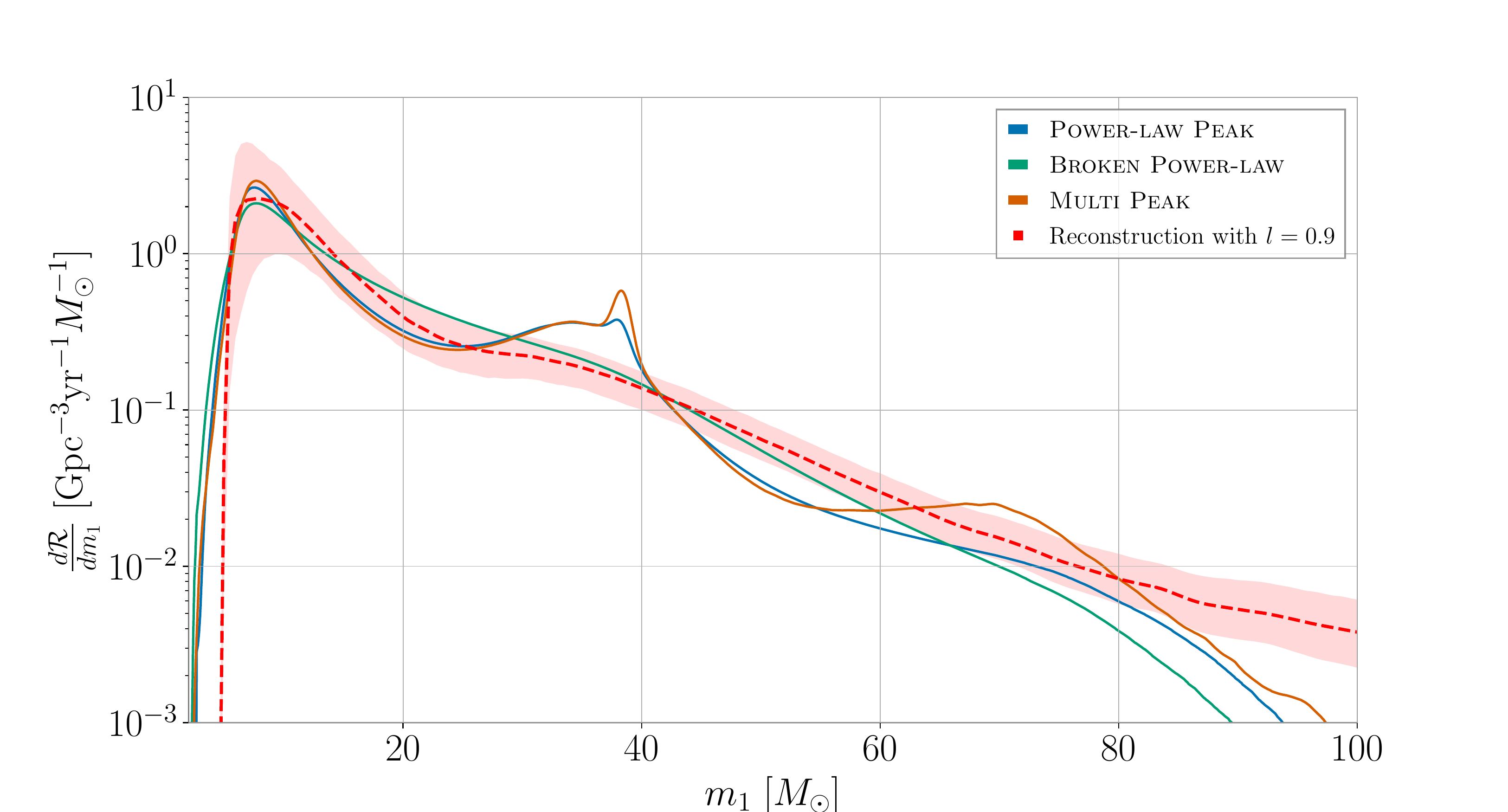}{0.8\textwidth}{}
    }
\caption{Distribution of primary mass obtained by the model with $l=0.9$ comparing to the models in \cite{2020arXiv201014533T}. The dashed red curve is the mean value over the posterior population distribution, the shaded region shows the 90\% credible interval, and the solid lines represent the mean value of distributions obtained by the 3 most favored models in \cite{2020arXiv201014533T}.}
\end{figure}
The reconstruction of component mass function from the inferred chirp mass distribution depends upon the distribution of mass ratio. As pointed out by \cite{2020arXiv201104502T}, there is no significant evidence for the mass ratio having a mass-dependent distribution. Therefore, we simply assume the distribution of mass ratio $p(q)\varpropto q^{1.4}$ as obtained in \cite{2020arXiv201014533T}, and additionally, the secondary mass is constrained to be $>2\,M_{\odot}$. Then the distribution of primary mass can be obtained (more details are presented in the Appendix \ref{rec_m1}). As expected, it has the feature similar to the distribution of chirp mass. Fig.~\ref{m1} shows the distribution of primary mass that converted from the chirp mass distribution (reconstructed with $l=0.9$). We notice that the astrophysical BBHs' primary mass function is not single-power-law-like either, and there must be some features presenting in the range of $30-40\,M_{\odot}$.
Four distributions reconstructed by \cite{2020arXiv201014533T} are also presented in Fig.~\ref{m1} for comparision. The peak in our model that located in $(30,\,40)\,M_{\odot}$ is lower than the peak of the \textsc{Power-law Peak} model \citep{2020arXiv201014533T}, while if comparing to the \textsc{Broken Power-law} model, there is a depression in the range of $(20,\,30)\,M_{\odot}$ for our model. In the high-mass range, there is no significant peak or cut off, this may be explained by the fact that we obtain the primary mass distribution from the reconstructed chirp mass function with a fixed mass ratio distribution.

Assuming that the merger rate density does not evolve with redshift, we integrate the BBH merger rates over the whole surveyed chirp mass range. This merger rate is then estimated to be $26.29^{+14.21}_{-8.96}\,\rm {Gpc^{-3}~yr^{-1}}$ for the model of $l=0.9$. The merger rates of BBH estimated with $l$ in the range of $(0.2,2.0)$ are found to be rather similar. 

\section{Summary and discussion}
We introduced a new method that rely on GP for reconstruction of the mass function of BBHs, and found that our method can well recover the injected distribution of simulated events. 
Comparing to the existing parametrization methods, our method is more faithful to the basic observation data, and it is able to distinguish the underlying features of mass distribution of BBHs when we know little about their formation processes.
We apply it to the real observed data, and obtained the chirp mass distributions reconstructed by models with different length scale $l$. By comparing the Bayes factors among models (with different $l$), we found that the chirp mass of a single-power-law-like distribution is disfavored. On the contrary, there must be a peak located in $(20,\,30)\,M_{\odot}$, and this may be contributed by the Population \uppercase\expandafter{\romannumeral3} BHs \citep{2021arXiv210300797K}. Meanwhile, there may be one or two peaks in the range of $< 20\,M_{\odot}$, which is consistent with the results of \citet{2020A&A...636A.104B}. Though there is no abrupt mass cut off in our surveyed chirp mass range,  the distribution function drops rather rapidly in the range of $30M_\odot<{\cal M}<60M_\odot$. Such a quick drop may be due to the superposition of the mass cutoff induced by the pair pulsational instability supernovae and the emergence of a more massive but sub-dominant population from for example the hierarchical mergers \citep[see][and the references therein]{2021ApJ...913...42W}.

In order to reconstruct the distribution of component masses, we assume a mass ratio distribution $p(q)\propto q^{1.4}$ as obtained in \cite{2020arXiv201014533T}, the distribution of primary mass turns out a same feature as that of chirp mass, i.e., there is a peak located in $(30,\,40)\,M_{\odot}$, this feature echo with those of \textsc{Power Law + Peak} and \textsc{Broken Power Law} in \cite{2020arXiv201014533T}.
Finally, we also calculate the merger rate of $26.29^{+14.21}_{-8.96}~\rm {Gpc^{-3}~yr^{-1}}$, which is consistent with that found in \cite{2020arXiv201014533T}.

Our current method is not able to model the mass ratio and spin properties of the BBH population together with chirp mass, and dedicated efforts to model these parameters are left to future work. Nevertheless, in the approach of a mixture of weighted Gaussians to reconstruct the chirp mass distribution together with spin parameters and mass ratio,  \citet{2020arXiv201104502T}  found that neither the mass ratio nor the aligned spin distributions shows significant dependence upon the mass. These authors also found out  two peaks locating at $7.1\,M_{\odot}$ and $13.6\,M_{\odot}$ respectively in the chirp mass function, which is consistent with some of our inferences (with small length scales, like $l=0.5$). However, whether these features are real or caused by the statistical fluctuation is still be clarified with more observations in the upcoming O4/O5 runs of the LIGO/Virgo/KAGRA detectors \citep[][; see also \url{https://dcc.ligo.org/public/0161/P1900218/002/SummaryForObservers.pdf}]{2018LRR....21....3A}.  

\acknowledgments

We thank the referee for the very helpful comments and suggestions. This work was supported in part by NSFC under grants of No. 11921003, No. 11933010, and No. 12073080, the Chinese Academy of Sciences via the Strategic Priority Research Program (Grant No. XDB23040000), Key Research Program of Frontier Sciences (No. QYZDJ-SSW-SYS024). This research has made use of data and software obtained from the Gravitational Wave Open Science Center (\url{https://www.gw-openscience.org}), a service of LIGO Laboratory, the LIGO Scientific Collaboration and the Virgo Collaboration. LIGO is funded by the U.S. National Science Foundation. Virgo is funded by the French Centre National de Recherche Scientifique (CNRS), the Italian Istituto Nazionale della Fisica Nucleare (INFN) and the Dutch Nikhef, with contributions by Polish and Hungarian institutes.\\

\vspace{5mm}
\software{Bilby \citep[version 0.5.5, ascl:1901.011, \url{https://git.ligo.org/lscsoft/bilby/}]{2019ascl.soft01011A},
          PyMultiNest \citep[version 2.6, ascl:1606.005, \url{https://github.com/JohannesBuchner/PyMultiNest}]{2016ascl.soft06005B},
          PyCBC \citep[gwastro/pycbc: PyCBC Release v1.14.1, \url{https://github.com/gwastro/pycbc/tree/v1.14.1}]{2019PASP..131b4503B,2019zndo...3546372N},
          gpflow \citep[\url{https://gpflow.readthedocs.io/en/master/intro.html}]{GPflow2017,GPflow2020multioutput}
          }

\bibliographystyle{aasjournal}
\bibliography{GPBHMF-bibtex}
   
\appendix

\section{reconstruction of primary mass}\label{rec_m1}
We reconstruct the distribution of primary mass from the chirp mass distribution and a mass-independent mass ratio distribution $p(q)\varpropto q^{1.4}$, additionally the secondary mass $m_2$ is constrained to be $>2\,M_{\odot}$. The detail of operation is as following.

For each posterior distribution of chirp mass $\boldsymbol{r}$ (that obtained in the Bayes inference), we first sample a sufficient number of samples $\{\mathcal{M}'_j\}$ from each bin, since the bins are narrow enough, it is reasonable to assume that the samples are uniformly distributed in each bin, the number of samples of $i$-th bin $N_i$ is proportional to the corresponding merger rate $r_i$ and width of bin $\Delta{\mathcal{M}}_{i}$ (i.e., $N_i \propto r_{i} \times \Delta{\mathcal{M}}_{i}$). For each chirp mass sample $\mathcal{M}'_j$, we sample a mass ratio $q'_j$ from power law: $\mathcal{P}(q|\alpha=1.4,q_{min},1)$, where $q_{min}$ can be calculated from $\mathcal{M}'_j$ and $m_2 > 2\,M_\odot$. Then we can obtain the samples of primary mass $\{{m_1}'_j\}$ from $\{(\mathcal{M}'_j,q'_j)\}$, and the distribution of primary mass can be obtained by Gaussian kernel density estimation (KDE).

\section{reconstruction with different length scales}\label{length}

\begin{figure}
\gridline{\fig{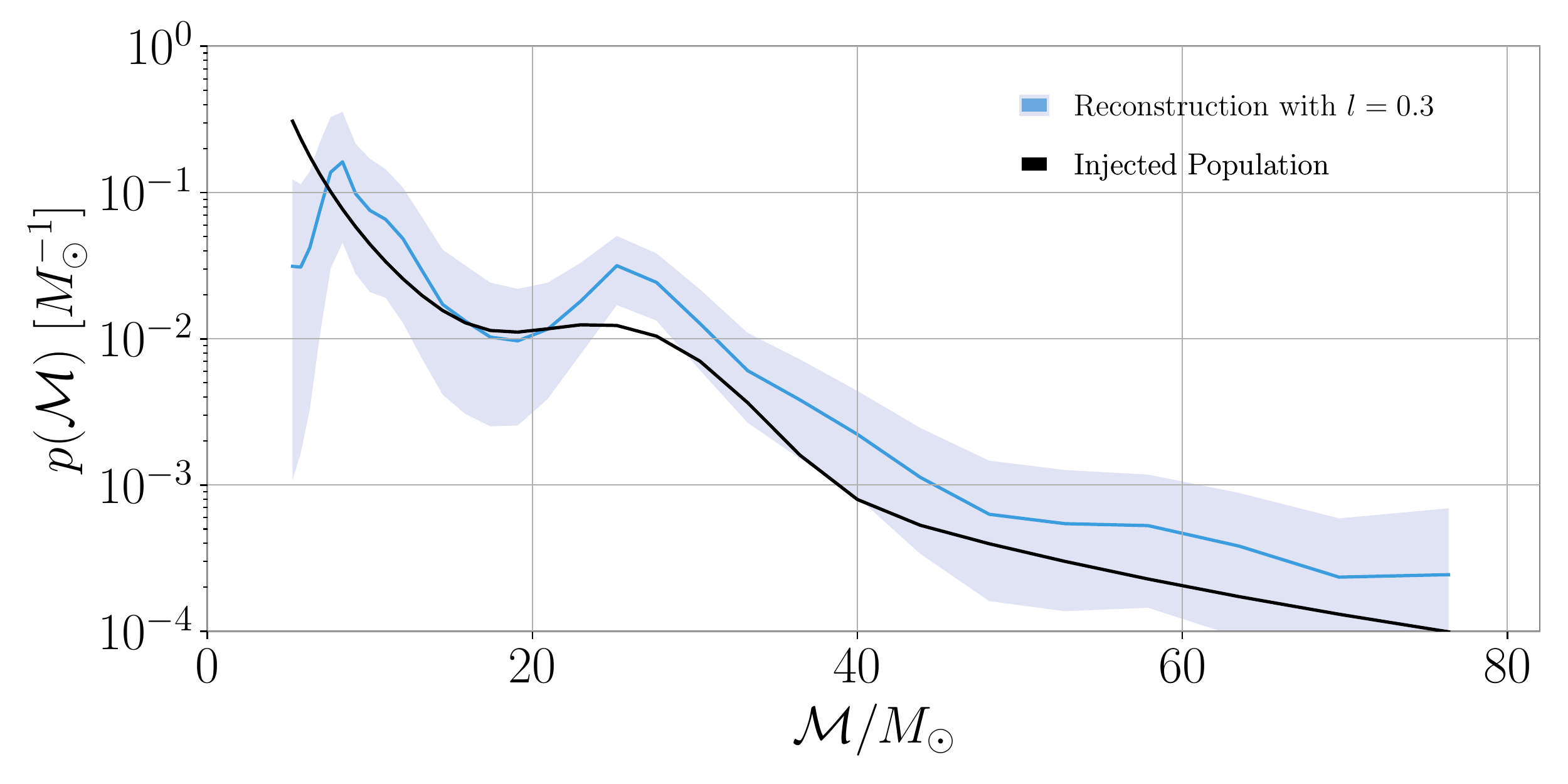}{0.33\textwidth}{}
\fig{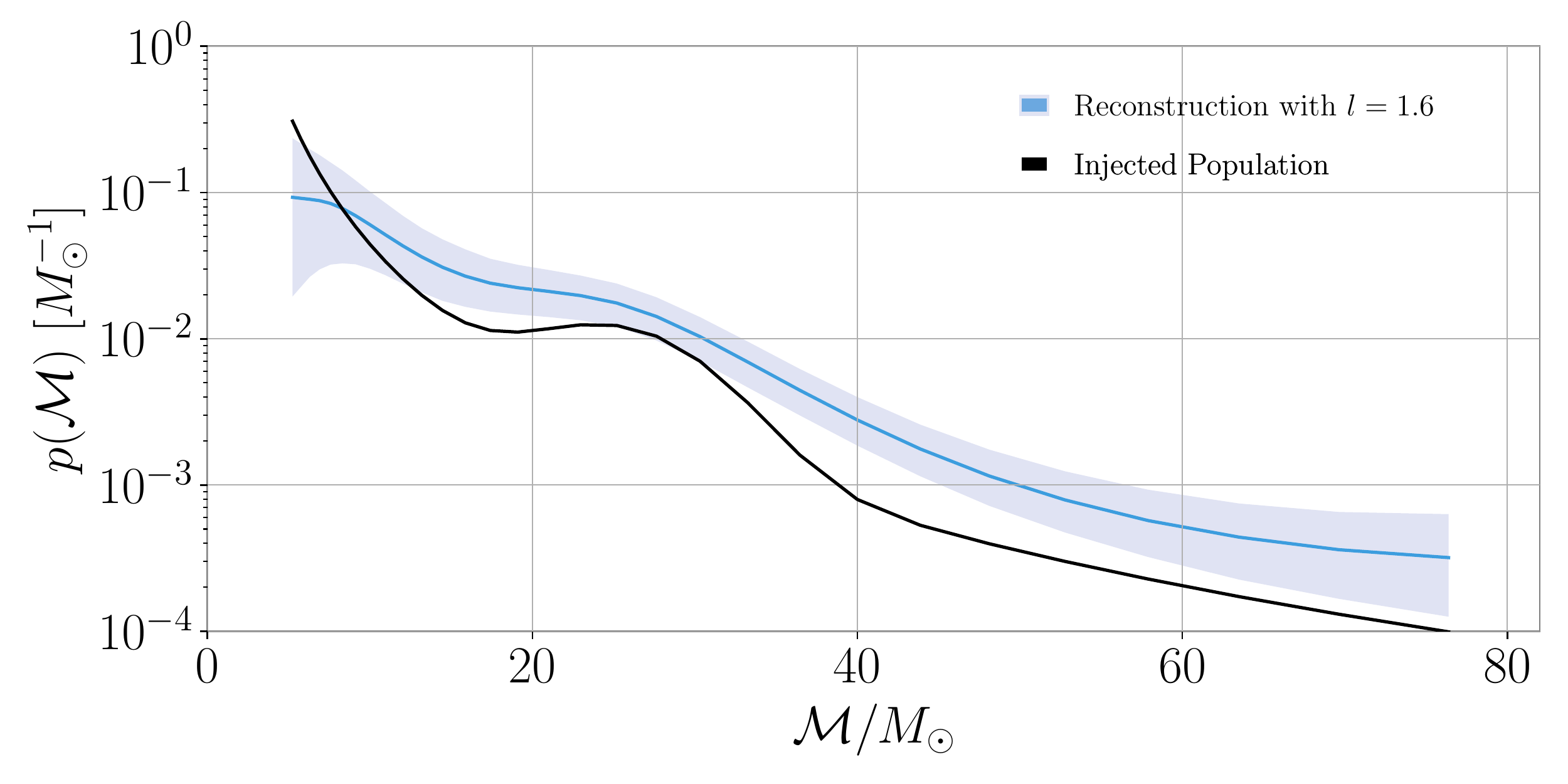}{0.33\textwidth}{}
\fig{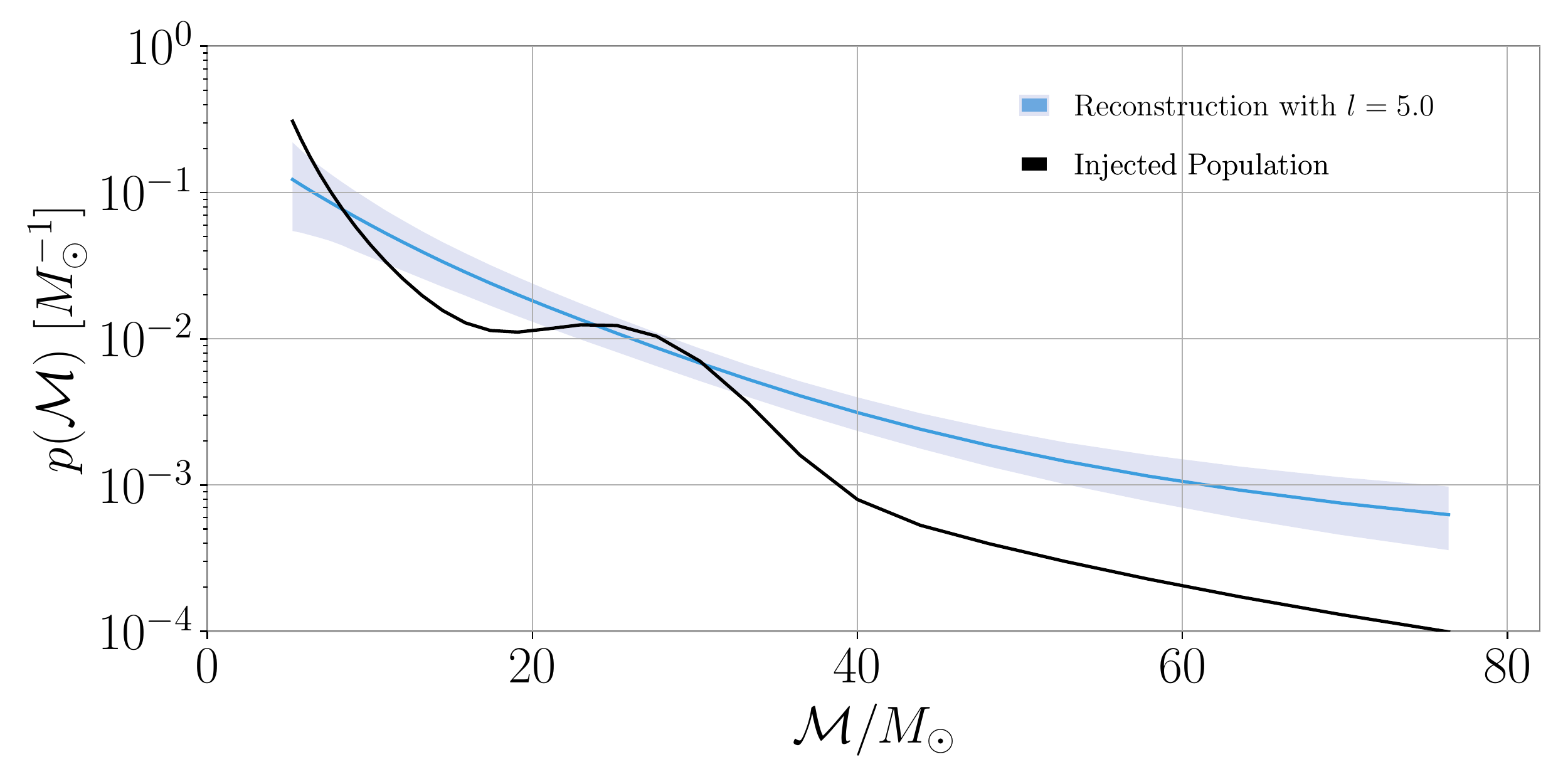}{0.33\textwidth}{}}
\gridline{\fig{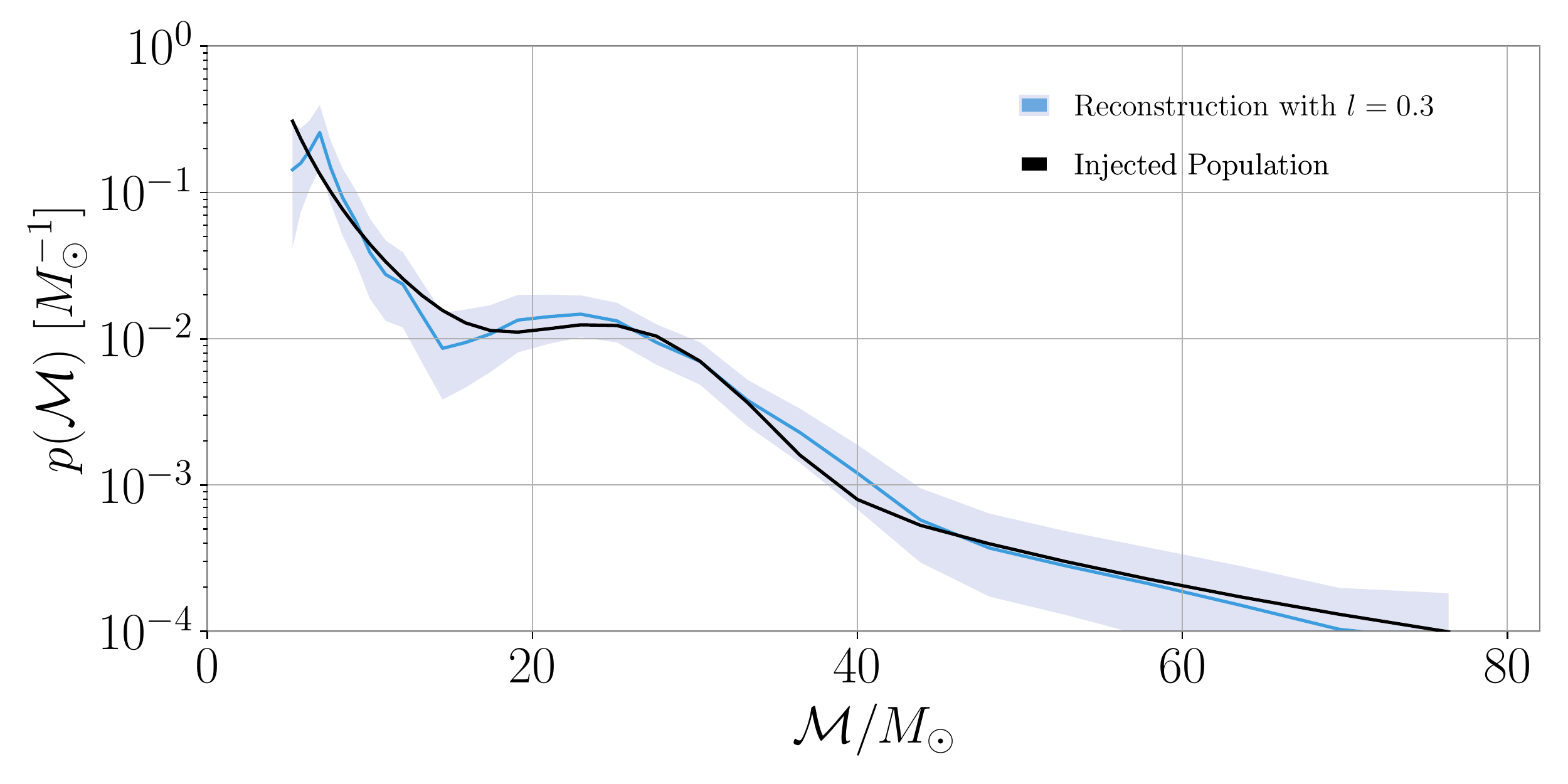}{0.33\textwidth}{}
\fig{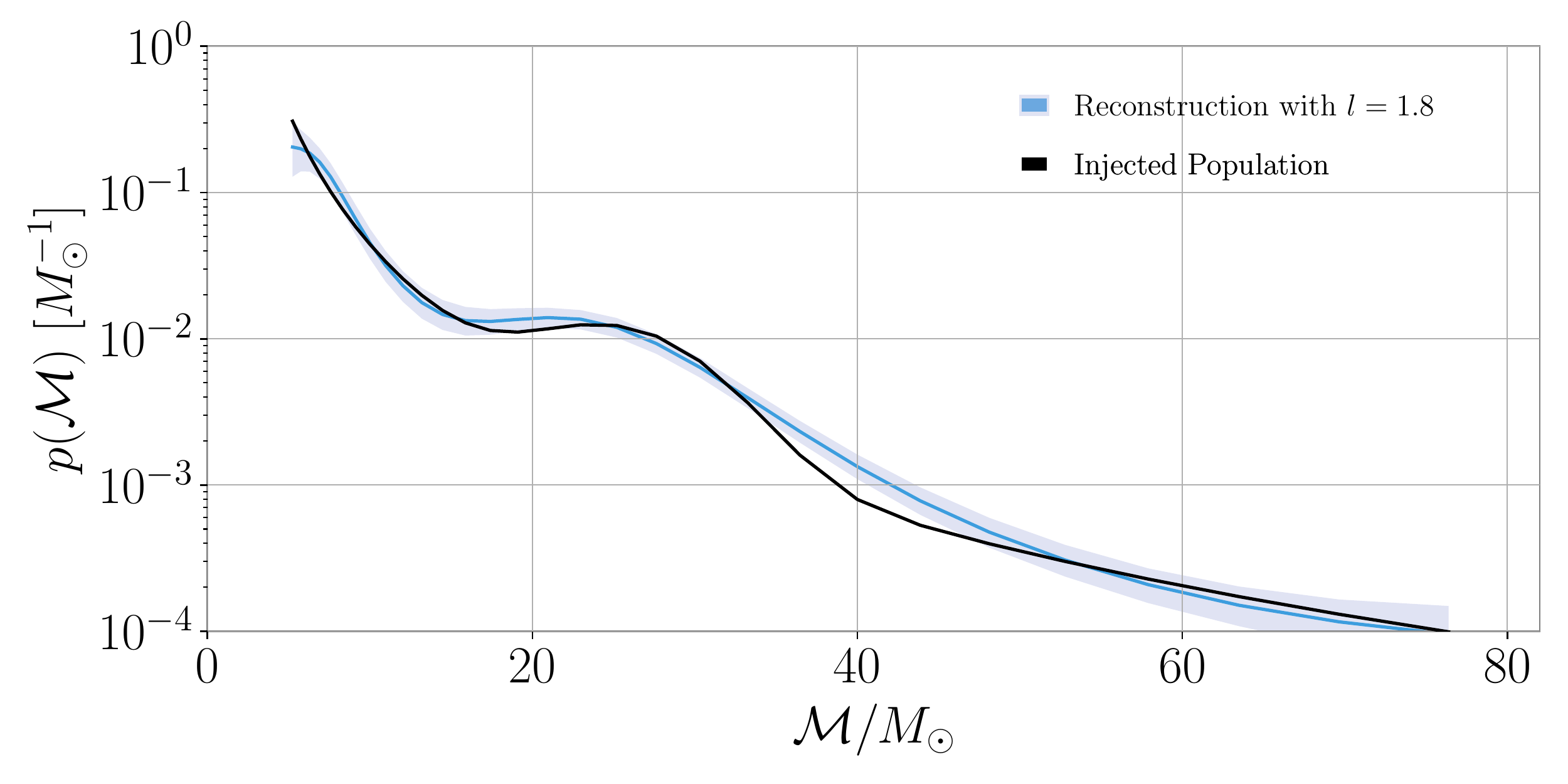}{0.33\textwidth}{}
\fig{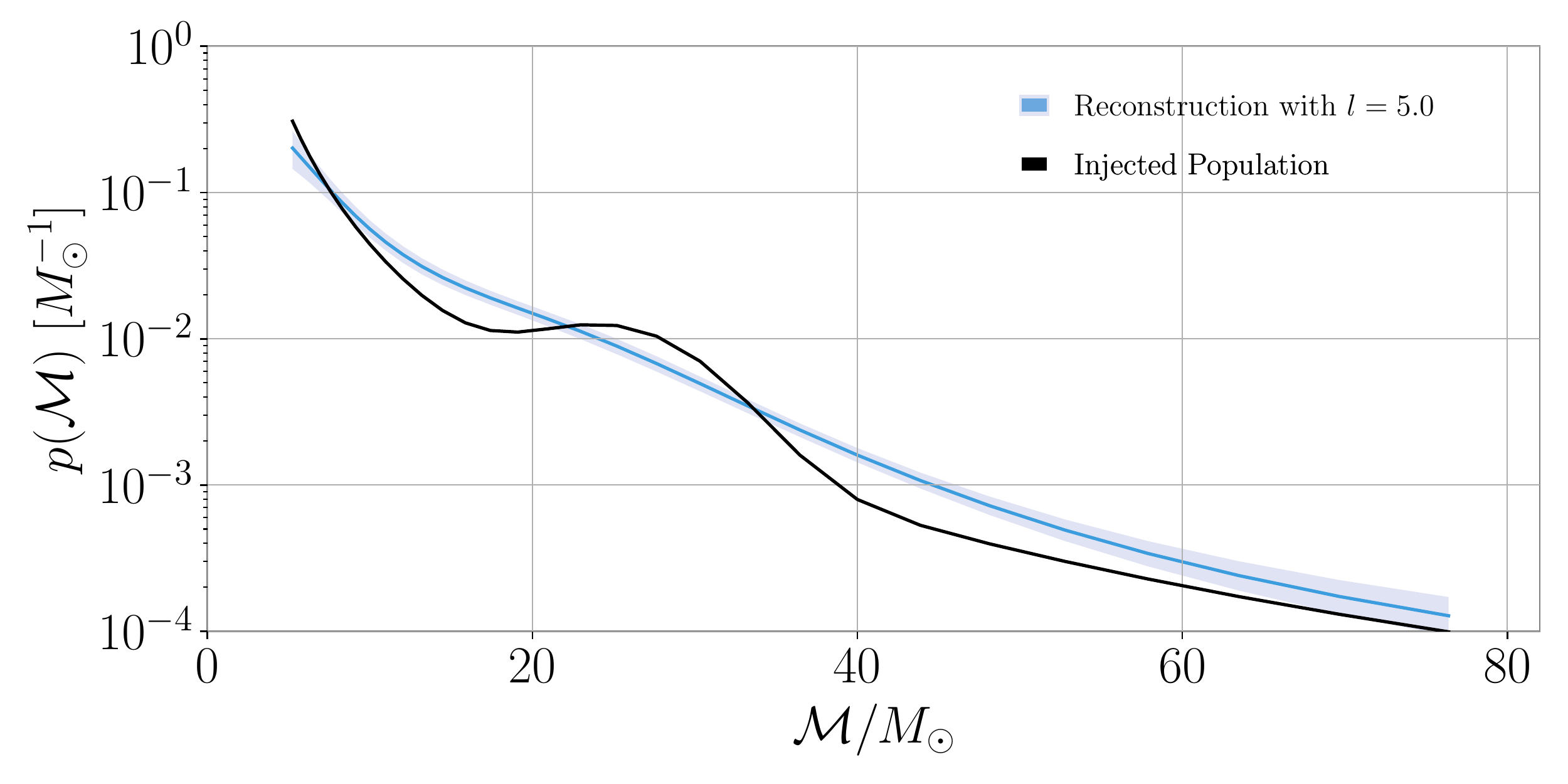}{0.33\textwidth}{}}
\gridline{\fig{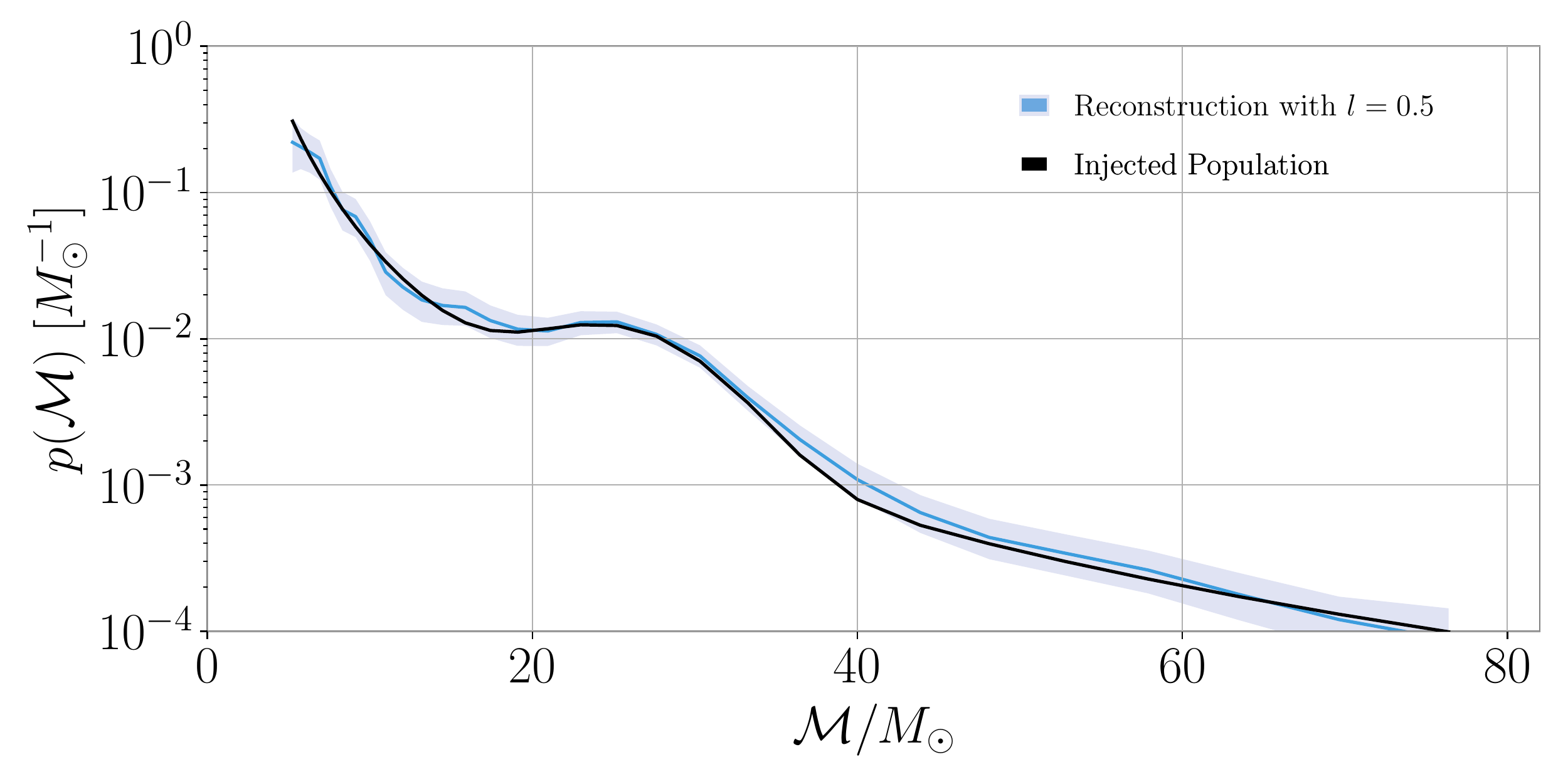}{0.33\textwidth}{}
\fig{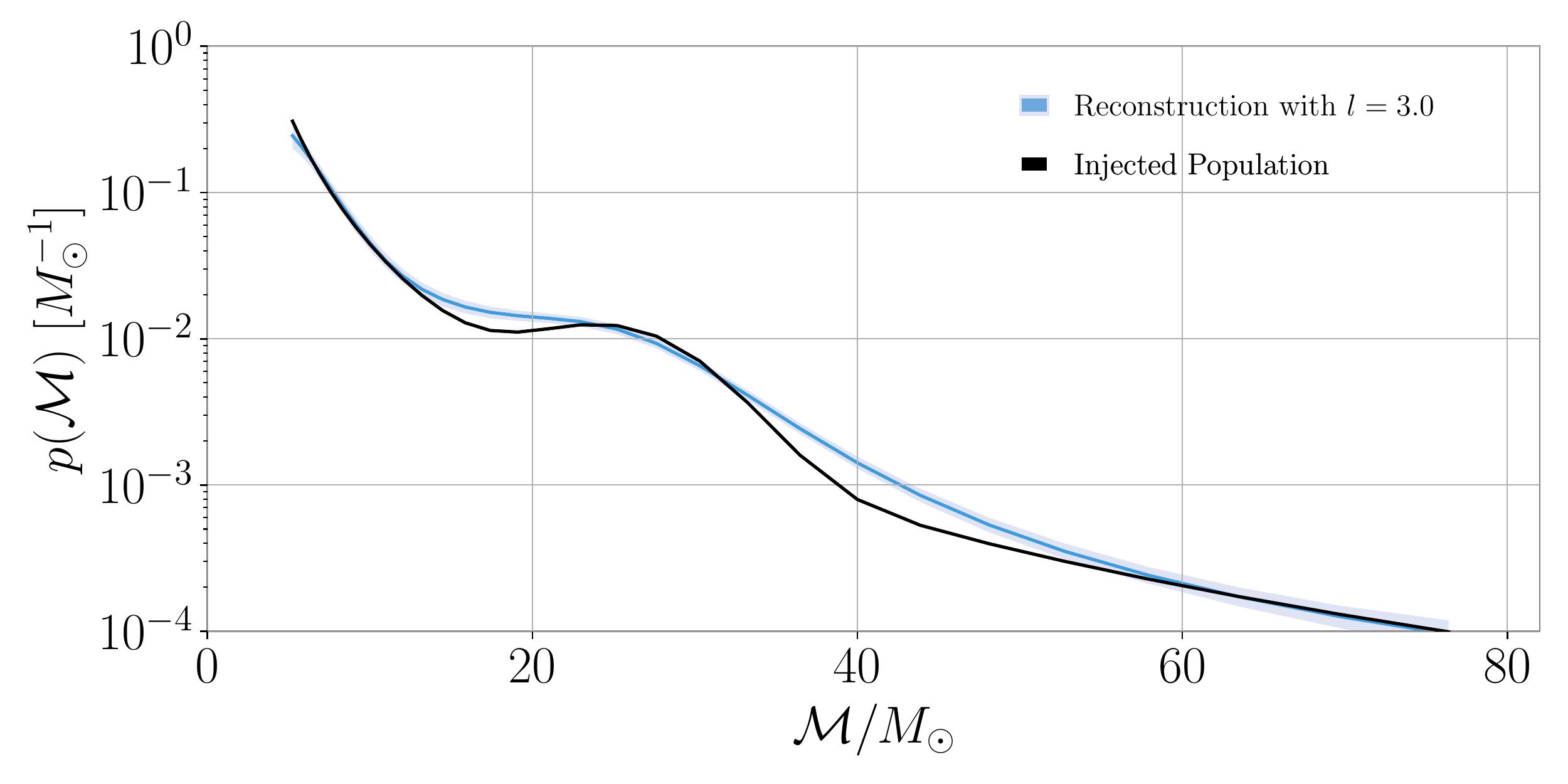}{0.33\textwidth}{}
\fig{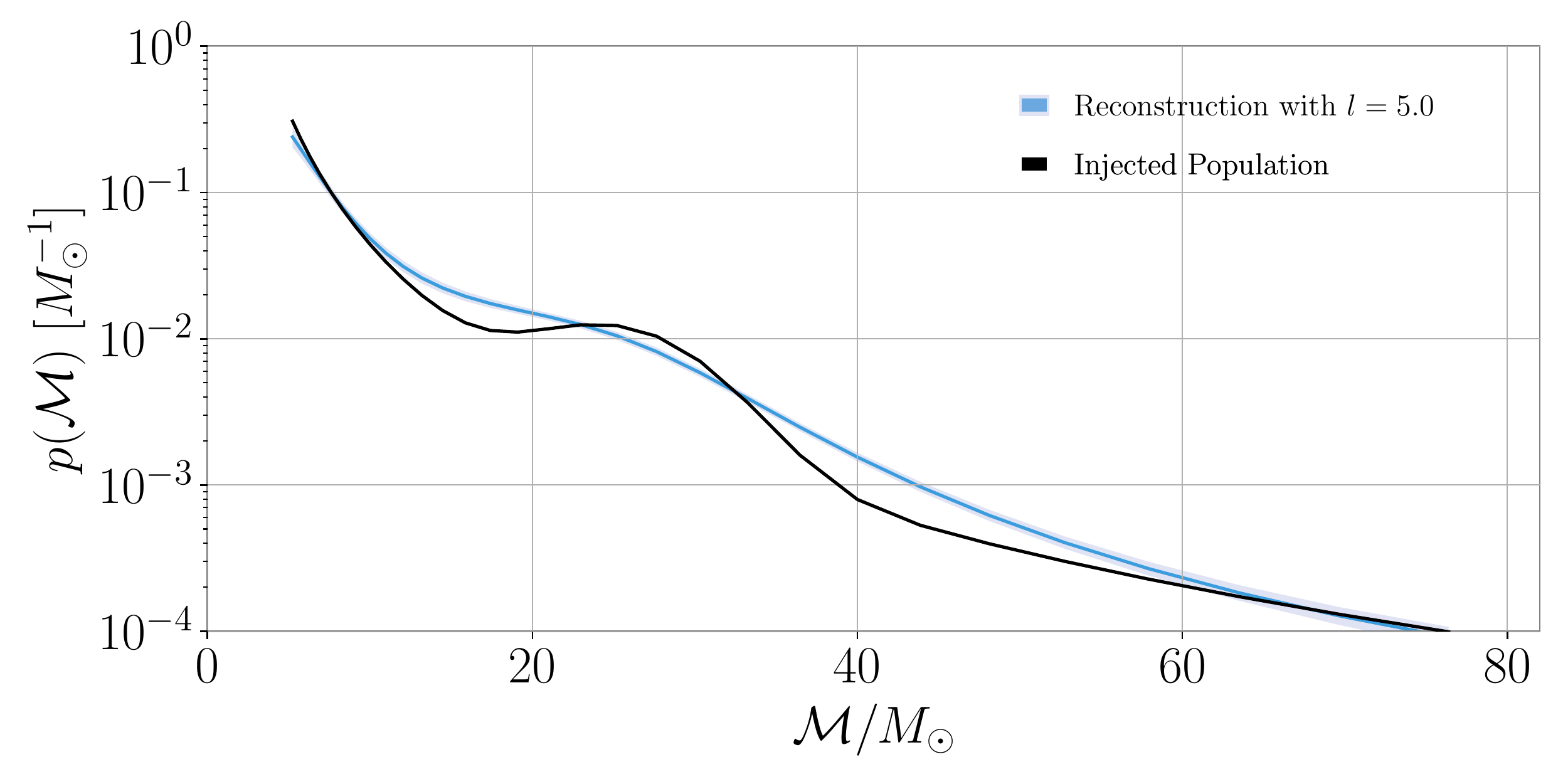}{0.33\textwidth}{}}
\caption{Normalized distributions of chirp mass reconstructed with several different length scales upon 40 (top), 300 (middle), and 1000 (bottom) simulated detections.}\label{sim_app}
\end{figure}

For each scenario of simulation, we display three reconstructions with different length scales (see Fig.~\ref{sim_app}). Referring to the corresponding Bayes factors in Fig.~(\ref{sim_rec}), we can rule out the single-power-law-like distribution, and identify the real structures (i.e., the peak at $25\,M_{\odot}$) in the injected distribution, additionally, we can rule out the fake features those are caused by the fluctuation of data as shown in the left column of Fig.~\ref{sim_app}. 

\begin{figure}
\gridline{\fig{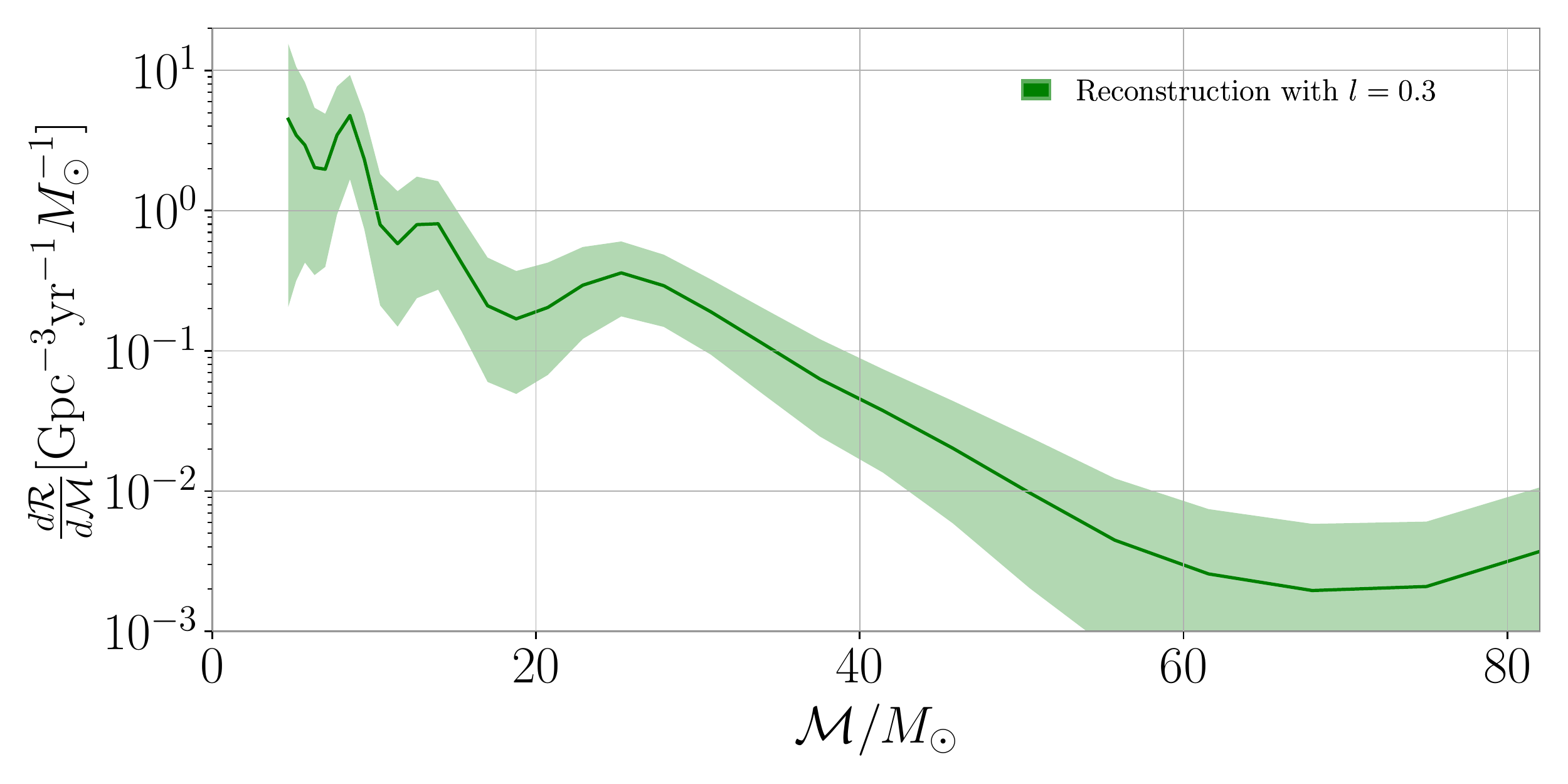}{0.5\textwidth}{}
\fig{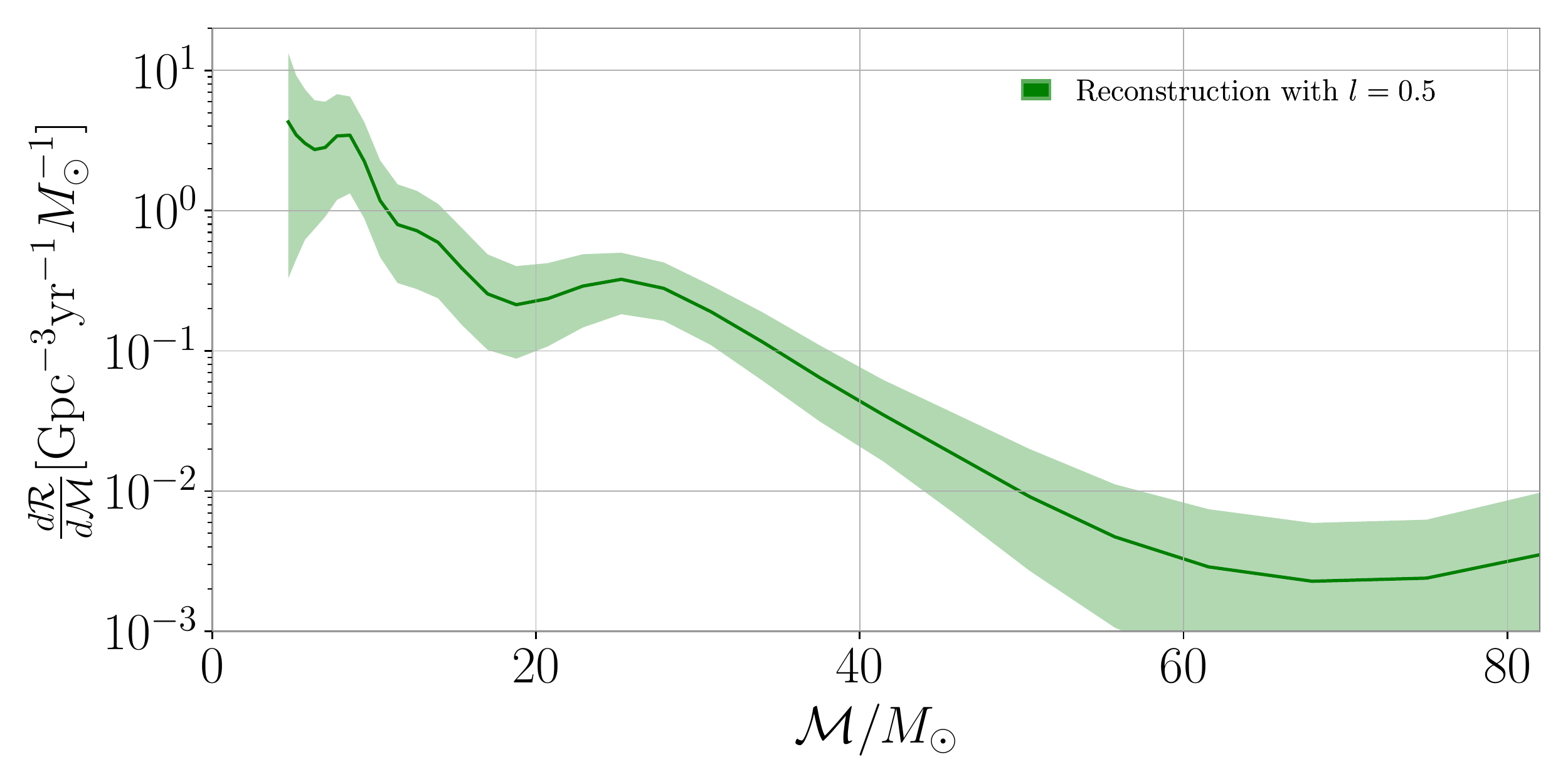}{0.5\textwidth}{}}
\gridline{\fig{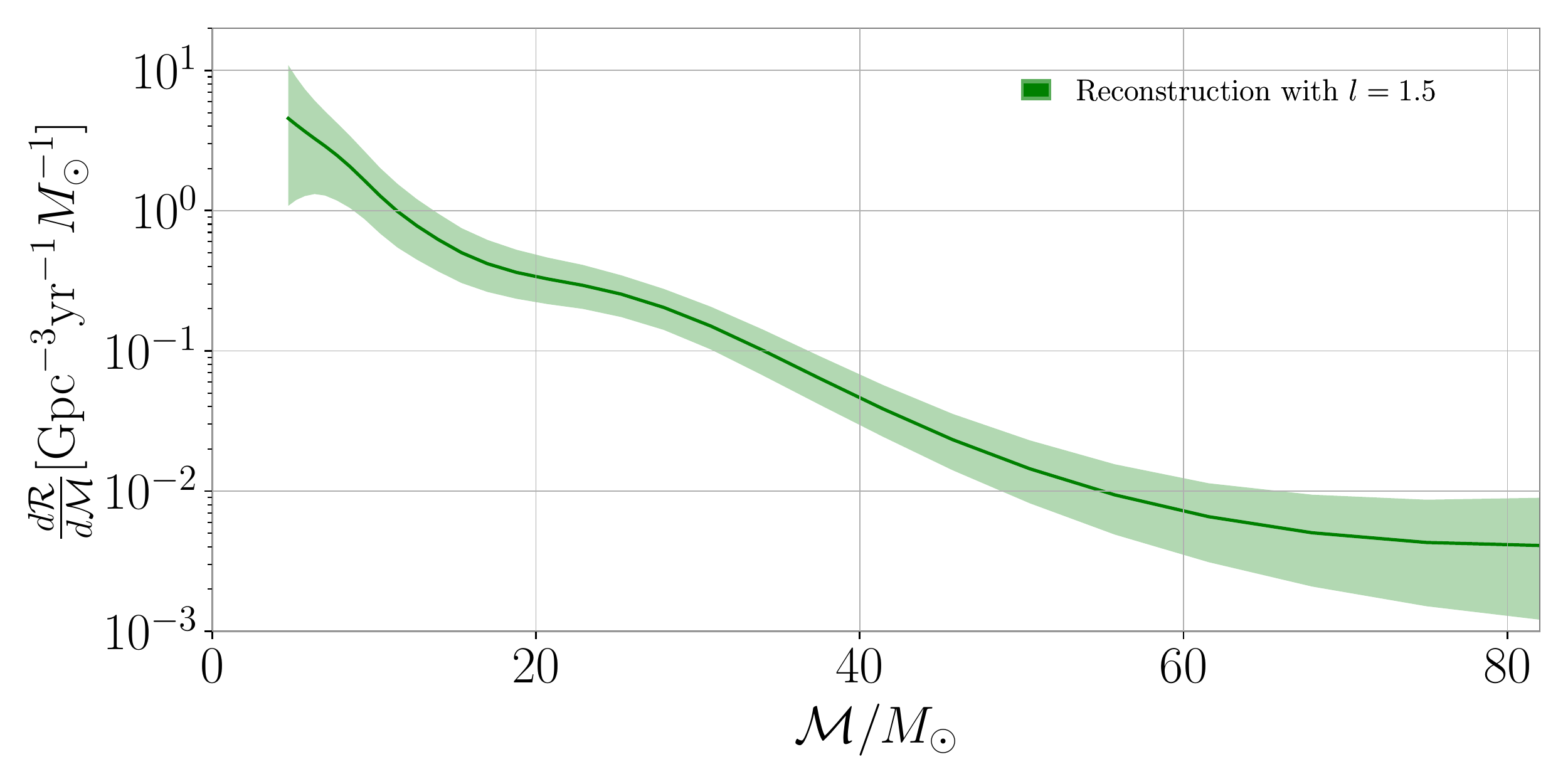}{0.5\textwidth}{}
\fig{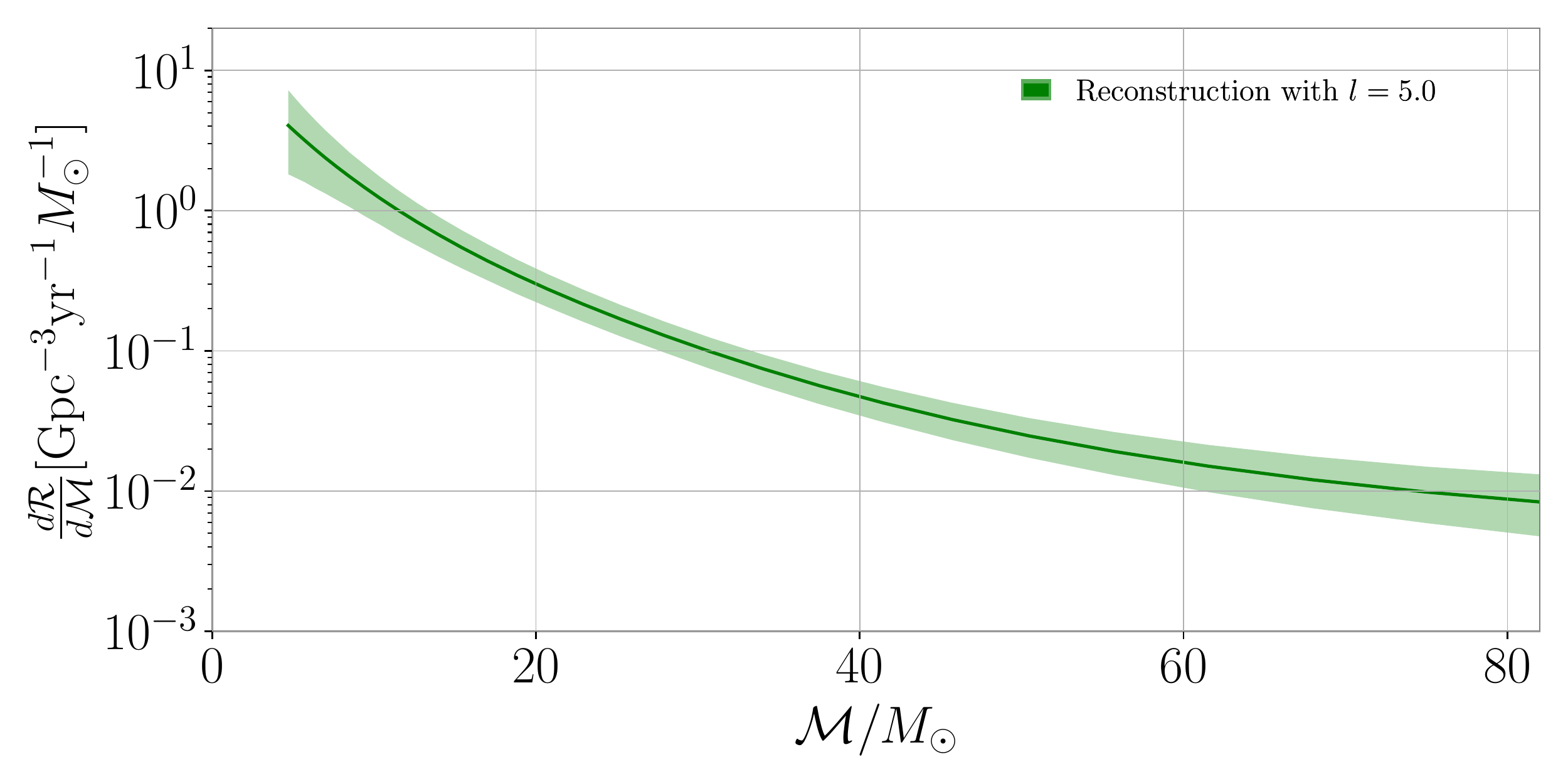}{0.5\textwidth}{}}
\caption{Distributions of chirp mass reconstructed with several different length scales upon real data.}\label{real_app}
\end{figure}

Fig.~\ref{real_app} shows the different reconstructions from real data. Considering their related Bayes factors in Fig.~\ref{Ls}, we can identify the peak in $(20,\,30)\,M_{\odot}$, and rule out the single-power-law-like distribution. While the two peaks in $(5,\,10)\,M_{\odot}$ and $(10,\,20)\,M_{\odot}$ can not be fully confirmed yet. 

\section{Result checking}
\begin{figure}[htbp]
\centering
\includegraphics[scale=0.5]{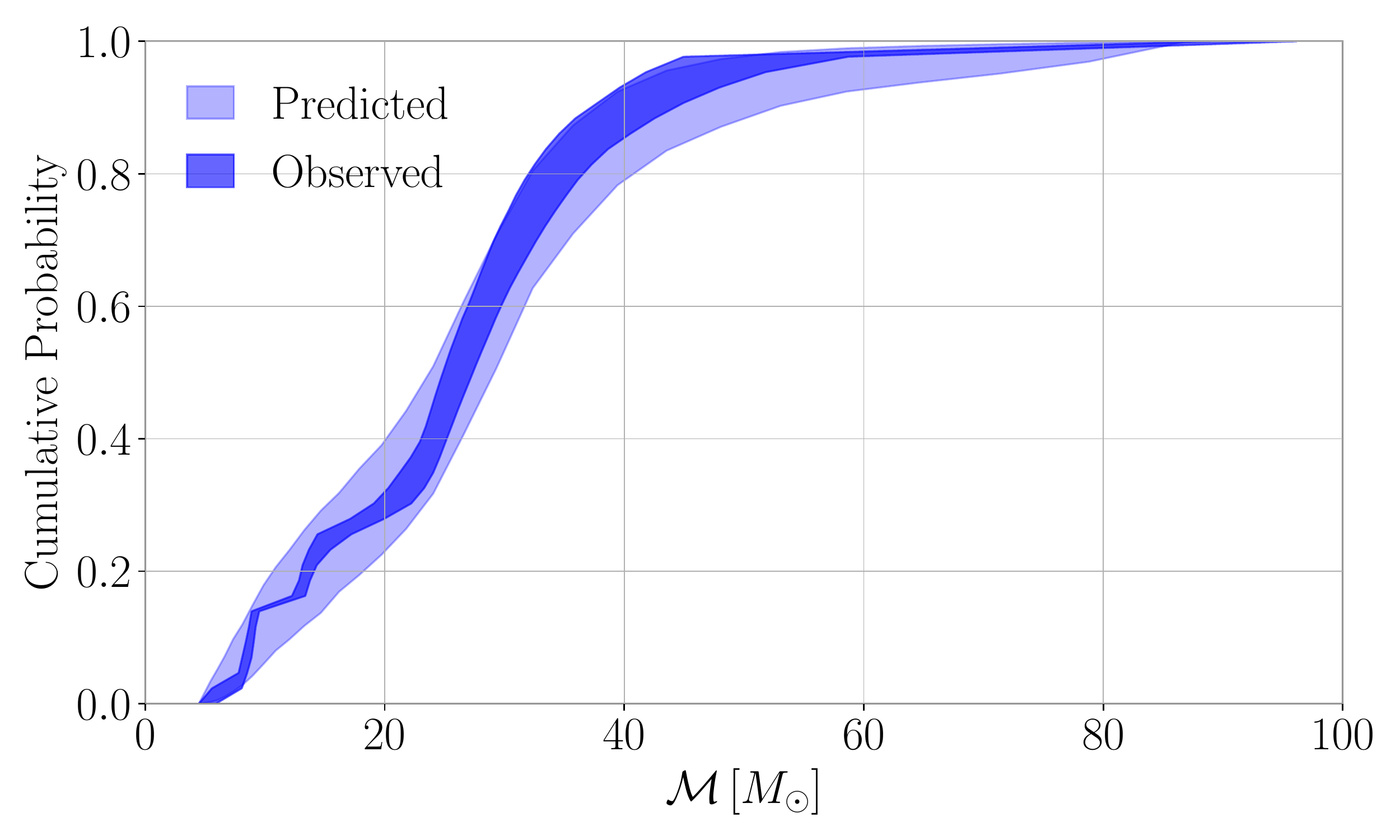}
\caption{A posterior predictive check: the cumulative density function (CDF) of the observed chirp mass distribution for the reconstruction with length scale $l=0.9$. The observed event distribution is shown in the darker colors. The thickness of the bands indicates the 90\% credibility range.}
\label{fig6}
\end{figure}

Section \ref{result} shows the inferred chirp mass distribution of BBHs with our method. Here we use the inferred distribution to predict an observed distribution, and compare it to the empirical distribution of observed chirp mass distributions in Fig.\ref{fig6}.
Note that empirical distribution is calculated from the re-weighted samples of the posteriors of the observed events (as introduced in section \ref{sec:llh}) rather than directly calculated from their posteriors. From Fig.~\ref{fig6} we can see that the dark colored band overlap with the light colored band in general, while our model dose not perform so well on the two edges of the mass range, this may be because that we fix the lower and upper bounds of the chirp mass (as 4.5$M_{\odot}$ and 87$M_{\odot}$) when fitting.

\end{CJK*}
\end{document}